\documentclass[preprint2]{aastex61}

\received{27 January 2019}
\accepted{\today}
\submitjournal{ApJ}

\usepackage{longtable}
\usepackage{mathrsfs}
\usepackage{url}
\usepackage[normalem]{ulem}
\usepackage{graphicx}

\usepackage{amsmath}	
\usepackage{amssymb}	

\usepackage{natbib}
\usepackage{color}
\bibpunct{(}{)}{;}{a}{}{,}

\newcommand\aproxgt{\mathrel{%
      \rlap{\raise 0.511ex \hbox{$>$}}{\lower 0.511ex \hbox{$\sim$}}}}
\newcommand\aproxlt{\mathrel{%
      \rlap{\raise 0.511ex \hbox{$<$}}{\lower 0.511ex \hbox{$\sim$}}}}

\newcommand{\xmm}{\it XMM-Newton}

\def\aap{AA}

\def\apjl{ApJ}
\def\apjs{ApJS}

\def\arcsec{\ifmmode {''} \else ${''}$\fi}
\def\arcmin{\ifmmode {'} \else ${'}$\fi}
\def\deg{\ifmmode {^\circ} \else ${^\circ}$\fi}
\def\cc{\ifmmode {\rm cm}^{-3} \else cm$^{-3}$\fi}
\def\cl{\ifmmode {\rm cm}^{-2} \else cm$^{-2}$\fi}
\def\pcm2{\ifmmode {\rm cm}^{-2} \else cm$^{-2}$\fi}
\def\micron{\ifmmode \mu{\rm m} \else $\mu$m\fi}
\def\kms{\ifmmode {\rm km\,s}^{-1} \else km\,s$^{-1}$\fi}
\def\kmps{\ifmmode {\rm km\,s}^{-1} \else km\,s$^{-1}$\fi}
\def\Hubble{\ifmmode {\rm km\,s}^{-1}\,{\rm Mpc}^{-1}
        \else km\,s$^{-1}$\,Mpc$^{-1}$\fi}
\def\ergsec{\ifmmode {\rm ergs\;s}^{-1} \else ergs s$^{-1}$\fi}
\def\ergcms{\ifmmode {\rm ergs\,cm}^{-2}\,{\rm s}^{-1}
          \else ergs\,cm$^{-2}$\,s$^{-1}$\fi}
\def\ergcmsA{\ifmmode {\rm ergs\,cm}^{-2}\,{\rm s}^{-1}\,{\rm \AA}^{-1}
          \else ergs\,cm$^{-2}$\,s$^{-1}$\,\AA$^{-1}$\fi}
\def\ergcmsHz{\ifmmode {\rm ergs\,cm}^{-2}\,{\rm s}^{-2}\,{\rm Hz}^{-1}
          \else ergs\,cm$^{-2}$\,s$^{-1}$\,Hz$^{-1}$\fi}
\def\Msun{\ifmmode M_{\odot} \else $M_{\odot}$\fi}
\def\Lsun{\ifmmode L_{\odot} \else $L_{\odot}$\fi}
\def\qo{\ifmmode q_{0} \else $q_{0}$\fi}
\def\Ho{\ifmmode H_{0} \else $H_{0}$\fi}
\def\ciii{C\,{\sc iii}}
\def\civ{C\,{\sc iv}}

\def\ciii{\ifmmode {\rm C}\,{\sc iii} \else C\,{\sc iii}\fi}
\def\civ{\ifmmode {\rm C}\,{\sc iv} \else C\,{\sc iv}\fi}
\def\cv{\ifmmode {\rm C}\,{\sc v} \else C\,{\sc v}\fi}
\def\cvi{\ifmmode {\rm C}\,{\sc vi} \else C\,{\sc vi}\fi}

\def\o5007{[O\,{\sc iii}]\,$\lambda5007$}

\def\six{Si\,{\sc x}}

\def\xmm{XMM-{\it Newton}}

\shorttitle{High velocity winds in Mrk 335}
\shortauthors{Longinotti  et al.}

\begin{document}


\title{The XMM-Newton/HST view of the obscuring outflow in the Seyfert Galaxy Mrk 335 observed at extremely low X-ray flux.}

\correspondingauthor{Anna Lia Longinotti}
\email{annalia@inaoep.mx}


\author[0000-0001-8825-3624]{Anna Lia Longinotti}
\affiliation{Instituto Nacional de Astrof\'isica, \'Optica y Electr\'onica, Luis E. Erro 1, Tonantzintla, Puebla, M\'exico, C.P. 72840}
\affiliation{CONACyT-INAOE}

\author[0000-0002-2180-8266]{Gerard  Kriss}
\affiliation{Space Telescope Science Institute, 3700 S. Martin Drive, Baltimore, MD 21218, USA}

\author[0000-0001-6291-5239]{Yair Krongold}
\affiliation{Instituto de Astronom\'ia, Universidad Nacional Aut\'onoma de M\'exico,  Circuito Exterior, Ciudad Universitaria, Ciudad de M\'exico 04510, M\'exico}

\author{Karla Z. Arellano-Cordova}
\affiliation{Instituto Nacional de Astrof\'isica, \'Optica y Electr\'onica, Luis E. Erro 1, Tonantzintla, Puebla, M\'exico, C.P. 72840}
\affiliation{Instituto de Astrof\'{\i}sica de Canarias, E-38200 La Laguna, Tenerife, Spain }
\affiliation{Departamento de Astrof\'{\i}sica, Universidad de La Laguna, E-38206 La Laguna, Tenerife, Spain}

\author[0000-0002-9214-4428]{Stefanie Komossa}
\affiliation{Max Planck Institut fuer Radioastronomie, Auf dem Huegel 69,53121 Bonn, Germany}

\author{Luigi Gallo}
\affiliation{Department of Astronomy and Physics, Saint Mary's University, Halifax, Canada}

\author[0000-0002-9961-3661]{Dirk Grupe}
\affiliation{Department of Earth and Space Sciences, Morehead State University, 235 Martindale Drive, Morehead, KY, USA}

\author{Smita Mathur}
\affiliation{Department of Astronomy, Ohio State University, 140 West 18th Avenue, Columbus, Ohio 43210-1173}

\author{Michael Parker}
\affiliation{ESAC, PO Box 78, 28691 Villanueva de la Ca\~nada, Madrid, Spain} 

\author{Anil Pradhan}
\affiliation{Department of Astronomy, Ohio State University, 140 West 18th Avenue, Columbus, Ohio 43210-1173}

\author{Dan Wilkins}
\affiliation{Kavli Institute for Particle Astrophysics and Cosmology, Stanford University, 452 Lomita Mall, Stanford, CA 94305, USA}

\begin{abstract}
The Seyfert Galaxy Mrk 335 is known for its frequent  changes of flux and spectral shape in the X-ray band  occurred during recent years. These variations may be explained by the onset of a wind that previous,  non-contemporaneous high-resolution spectroscopy in X-ray and UV bands located at accretion disc scale. A simultaneous new campaign by {\it XMM-Newton} and {\it HST} caught the source at an historical low flux in the X-ray band. The soft X-ray spectrum is dominated by prominent emission features, and by the effect of a strong ionized absorber with an outflow velocity of 5-6$\times$10$^3$~km~s$^{-1}$. The broadband spectrum obtained by the EPIC-pn camera reveals the presence of an additional layer of absorption by gas at moderate ionization covering $\sim$80$\%$ of the central source, and tantalizing evidence for absorption in the Fe~K band  outflowing at the same velocity of the soft X-ray absorber. The {\it HST-COS} spectra confirm the simultaneous presence of broad absorption troughs in CIV, Ly$\alpha$, Ly$\beta$ and OVI, with velocities of the order of 5000~km~s$^{-1}$ and covering factors in the range of 20-30\%.  Comparison of the ionic column densities and of other outflow parameters in the two bands show that the X-ray and UV absorbers are likely originated by the same gas.   The resulting picture from this latest multi-wavelength campaign confirms that Mrk~335 undergoes the effect of a patchy, medium-velocity outflowing gas in a wide range of ionization states  that seem to be persistently obscuring the nuclear continuum.  
\end{abstract}
\keywords{Active Galaxies: general --- Active Galaxies: Mrk 335}

\section{Introduction}
\label{sec:intro}
Accretion onto supermassive black holes is commonly regarded as one of the most distinctive feature of Active Galaxies (AGN). 
Nonetheless, the apparently counteractive process of ejection has been gaining growing importance in our knowledge of the AGN phenomenology thanks to the many observational results that have emerged in the last two decades. Ejection of gas in the form of winds is now a common property of radio-quiet AGN and it is particularly prominent in  X-ray and UV spectra of local sources (see \citet{Crenshaw03} for a review). Association of the X-ray winds to absorption lines detected in the ultraviolet band was historically proposed to relate the properties of the gas in the two bands possibly indicating a common origin \citep{Mathur95, Mathur98}.
However, the origin of the winds is not uniquely determined in all AGN where this phenomenon is observed: at parsec-scale thermal outflows arise from the molecular torus \citep{KK01}, whereas the accretion disc can launch outflows at a radial distance lower than 10$^3$ gravitational radii via radiative \citep{Proga04} and magneto-hydrodynamical mechanisms \citep{Konigl94,Fukumura2010}.

In the X-ray domain, ionization state, velocity and column density of the outflowing gas are relatively easy to measure. They provide information on the physical properties and, partly, on the location of the outflows although the radial distance cannot be measured directly in the spectra unless variability of the ionizing continuum takes place \citep[e.g.][]{Krongold07}. 
When the distance of the wind is sufficiently well pinned down, from the velocity and the column density it is possible to estimate the mass and energy output expelled by the outflow and quantify the impact that gas ejection may have onto the host galaxy. This is particularly important to understand the role of AGN winds in feedback process  (see \citet{King15} for a review). 
The multitude of results obtained via X-ray spectroscopy of bright AGN have shown that ``slow" outflows of ionized gas, widely known as ``warm absorbers", do not reach the minimum energy output required to alter the AGN star formation activity (e.g. \cite{Krongold07,Krongold10}). Instead, the so-called Ultra Fast Outflows (UFO), which are significantly faster and more massive, were shown to be capable of triggering  AGN feedback \citep{Tombesi15,Feruglio15,Longinotti18}. 

In the very recent years, another flavour of AGN winds is being observed and studied: with outflow velocity half-way between warm absorbers and UFOs, obscuring and intermittent outflows produced by clouds orbiting in the Broad Line Region (BLR) are now  revealed in the X-ray and UV spectra of Seyfert Galaxies that undergo important flux and spectral variations on relatively short time scales. Mrk~335 \citep{Longinotti13}, NGC5548 \citep{Kaastra14}, NGC~985 \citep{Ebrero16}, NGC3783 \citep{Mehdipour17,Kriss2018} are some examples of obscuration produced by intervening ionized gas in form of  ``eclipsing" winds.

The close-by Narrow Line Seyfert 1 Galaxy  Mrk~335 \citep[{\it z} = 0.025785,][]{Huchra99} is one of the few sources where the emergence of an obscuring wind outflowing at a velocity of 5-6000~km~s$^{-1}$ was revealed in the X-ray and UV bands \citep{Longinotti13}, based on non-simultaneous observations obtained by {\it XMM-Newton} and the {\it Hubble Space Telescope}. This work also provided the first record of X-ray ionized absorption seen in gratings spectra in Mrk~335, a source that in its past UV and X-ray history had showed little evidence for the presence of a typical warm absorber. 

Previous works based on CCD-resolution X-ray spectra of Mrk 335 were focussed to model in great detail the spectral curvature in terms of  intervening gas partially covering the line of sight, and/or relativistically
 blurred reflection from the accretion disk  \cite[e.g.][]{Turner93,Grupe2007,Grupe08,Larsson08,Grupe12,Gallo2013,Gallo2015,Wilkins15,Gallo2019}.

Indeed, in the past Mrk~335 was mostly known as a typical bright Seyfert 1 Galaxy \cite[e.g.][]{Longinotti2007,O'Neill07} with relativistic Fe K features and negligible intervening absorption until  the year 2007 when {\it Swift} discovered it in a very low X-ray  flux state \citep{Grupe2007}.  The ongoing monitoring with {\it Swift} since the sudden drop in 2007 has shown that Mrk~335 has remained in this dim X-ray  state with some occasional variability detected  along the elapsed  $\sim$11~years  \citep{Gallo2018}.  
According to the monitoring reported  by these authors, the long dimmed X-ray state  has not been accompanied by  corresponding variability in the Optical/UV band, which on average stays similar to measurements obtained prior to 2007. 
On the contrary, repeated X-ray flaring and dipping episodes have triggered several deep follow-up observations with XMM-Newton, Suzaku and NuSTAR  \cite[e.g.][and references above]{Parker2014,Gallo2015,WilkGal15,Keek16,Komossa17}.

In the present  paper the outcome of a more recent simultaneous campaign performed on Mrk~335 by {\xmm} and {\it HST} is presented. 
The chief goal of this campaign was to determine the properties of the absorbers with contemporaneous X-ray and UV data. 

The {\xmm} and HST observations of Mrk~335 were triggered in December 2015 following a decrease of the UV and X-ray flux revealed by {\it Swift} \citep{Grupe15}. 
{\xmm} started observing on  December the 30h for a total duration of $\sim$140 ks  (OBSID 0741280201) and HST followed on 4 and 7 of January 2016 for a total of 7 orbits. 
Unlike the previous report  \citep{Longinotti13} that relied on archival data obtained 4 months apart,  the above timing provides quasi-simultaneity between the  observational properties in UV and X-ray bands.  

\section{XMM-Newton observation and spectral analysis}
\begin{figure}[t]
	\includegraphics[width=0.85\columnwidth,angle=-90]{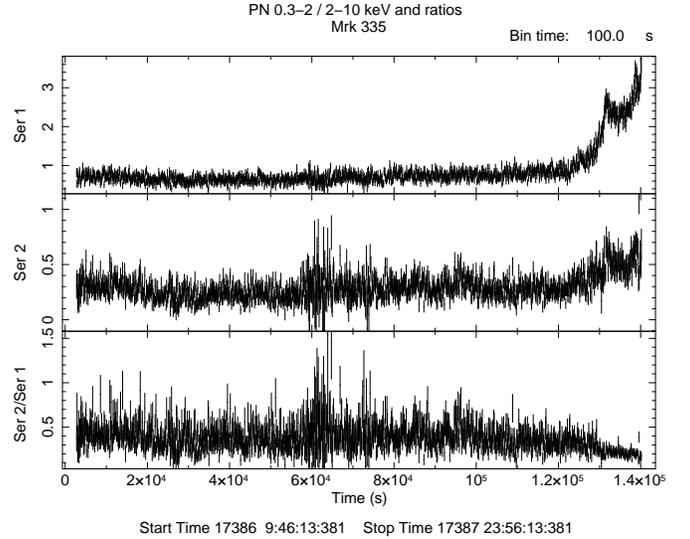}
    \caption{Background-subtracted light curve of Mrk~335 during the XMM-Newton observation. From top to bottom: source counts in 0.3-2 keV, 2-10 keV and  hardness ratio. The flux rises in the last 20~ks of the observation.}
    \label{fig:pn_lc}
\end{figure}

The {\xmm} EPIC cameras were both set to operate in Full Window mode. Data were processed with SAS 16.0.0\footnote{\tt https://www.cosmos.esa.int/web/xmm-newton/sas}. Both pn and MOS detectors were exposed to strong protons flares during some portions of the light curve. We followed the procedure recommended in the Science Threads to clean the raw event file and we obtained a clean exposure of 115~ks in the pn camera. 
Data obtained by the RGS instrument were not affected by the flaring periods.
Spectra were produced by standard SAS tools ({\tt rgsproc, epproc, emproc}). EPIC source and background counts were extracted from regions of 40~arcsec of radius, including all patterns (0-4 for the pn and 0-12 for the MOS cameras). 
After checking consistency among EPIC spectral products, the analysis of MOS spectra was not included in this paper.

\begin{figure}[t]
	\includegraphics[width=0.7\columnwidth,angle=-90]{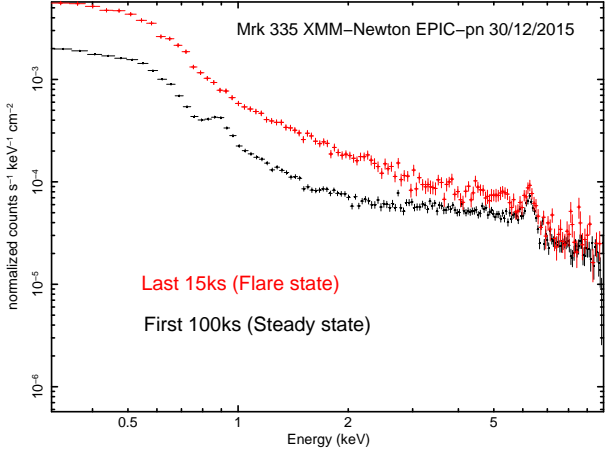}
    \caption{EPIC pn spectra extracted from the steady part of the light curve (first $\sim$100ks of the observation)  and from the "flare" (last 15~ks). The Fe~K band is moderately affected by the spectral change induced by the rise of flux. Both spectra  are severely affected  by  high background above 8~keV.}
    \label{fig:pn_spec}
\end{figure}
The light curve extracted from the pn counts shows an increase of the flux of a factor of $\sim$3 that happened during the last $\sim$20 ks of the observation (Fig.~\ref{fig:pn_lc})
We therefore split the data into ``steady" (first $\sim$120~ks) and ``flare" (last $\sim$20~ks) states to check for possible spectral variations. The spectral analysis described in the following was applied consistently to both flux states and model parameters were compared. 
Due to the combined effect of the source being at a very low flux state and of the short exposure in the spectral products extracted from the flaring portion of the light curve  flare (20~ks for the RGS and 15ks for the pn), no significant variation was detected in the spectral parameters corresponding to the two flux states. 

\begin{figure*}[t]
\hspace{-1.5cm}
		\includegraphics[width=8cm,height=18cm,angle=-90]{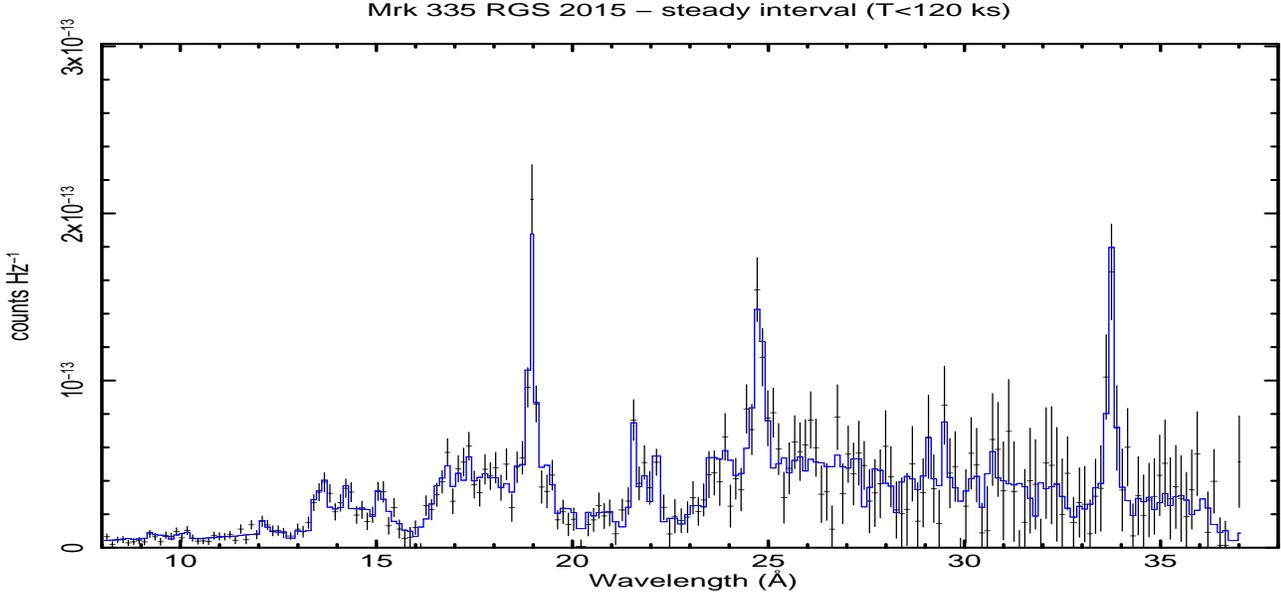}
    \caption{Best fit of the RGS spectrum corresponding to the steady interval of the light curve.  The continuum is modeled with a power law with $\Gamma$=2.72, the model includes the emission features and the ionized absorber described in Section~\ref{sec:rgs}. Data have been binned for plotting purpose. }
    \label{fig:rgs_spec}
\end{figure*}

However, as extensively described in the following section, the spectrum extracted from the ``steady" interval of the light curve is completely dominated by emission lines (see Fig.~\ref{fig:rgs_spec} and Table~\ref{tab_linelist}) that are not entirely recovered in the spectrum extracted from the ``flare" because the continuum emission starts covering them as the flux rises. This effect can be visualized  in Fig.~\ref{fig:pn_spec}. 
 With the aim of obtaining an accurate characterization of the soft X-ray features,  the spectral analysis was applied to the ``steady" interval ($\sim$120~ks for the RGS and $\sim$100ks for the pn). After obtaining a sound characterization of the underlying soft X-ray features a general consistency of this spectrum with the the ``flare"  interval was confirmed. Since the inclusion of the  ``flare"  interval does not provide any tighter constraints on our spectral analysis,  we decided not to include it in the present analysis and to focus on the results obtained on the ``steady" interval, which represent the most genuine description of the lowest X-ray spectrum of Mrk~335 so far obtained. The detailed analysis of the  ``flare"  interval is reported by \cite{Gallo2019}. 


The spectral analysis was carried out with XSPEC version 12.9.1 \citep{Arnaud96}. 

The analysis of the grating spectrum was carried out on the combined RGS1+RGS2 data sets obtained through the SAS tool {\tt rgscombine}.
Counts were not binned, therefore the Cash Statistics was applied \citep{Cash79}.
The EPIC-pn spectrum was binned by the SAS tool {\tt specgroup} therefore $\chi^2$ statistics was applied throughout its analysis  (see Section~\ref{sec:broadband})
Error bars of spectral parameters are quoted to 1$\sigma$. 
 Fluxes in the 0.3-10~keV and in the 0.3-2~keV bands during the steady part of the {\xmm} observation were measured to be $\sim$3.7$\times$10$^{-12}$  and $\sim$1.1$\times$10$^{-12}$ erg cm$^{-2}$ s$^{-1}$ respectively. These fluxes translate to X-ray luminosity of L$_{0.3-10}$=5.5$\times$10$^{42}$~erg~s$^{-1}$ and  L$_{0.3-2}$=1.5$\times$10$^{42}$~erg~s$^{-1}$. 

\subsection{The high resolution X-ray spectrum}
\label{sec:rgs}

The entire RGS spectrum is plotted in Figure~\ref{fig:rgs_spec}.  
The spectrum is dominated by very intense emission lines and Radiative Recombination Continua (RRC) features, along with strong signature of ionized absorption. The emission component was discovered in a previous low-flux, shorter observation  by {\xmm} in 2007 \citep{Grupe08} and reported by \cite{Longinotti08}. The higher signal-to-noise of the present data allows the effect of absorption lines to be clearly distinguished in the continuum emission.  This is not surprising since in another previous {\xmm} observation obtained in 2009 the emergence of a strong multi-layer ionized absorber was revealed in both X-rays and UV bands  \citep{Longinotti13}. 

The latest (2015) RGS view of Mrk 335 shows clear signatures of both features, therefore we constructed a model that includes emission and absorption  guided  by our previous knowledge of the spectrum. 
\begin{figure*}[t]
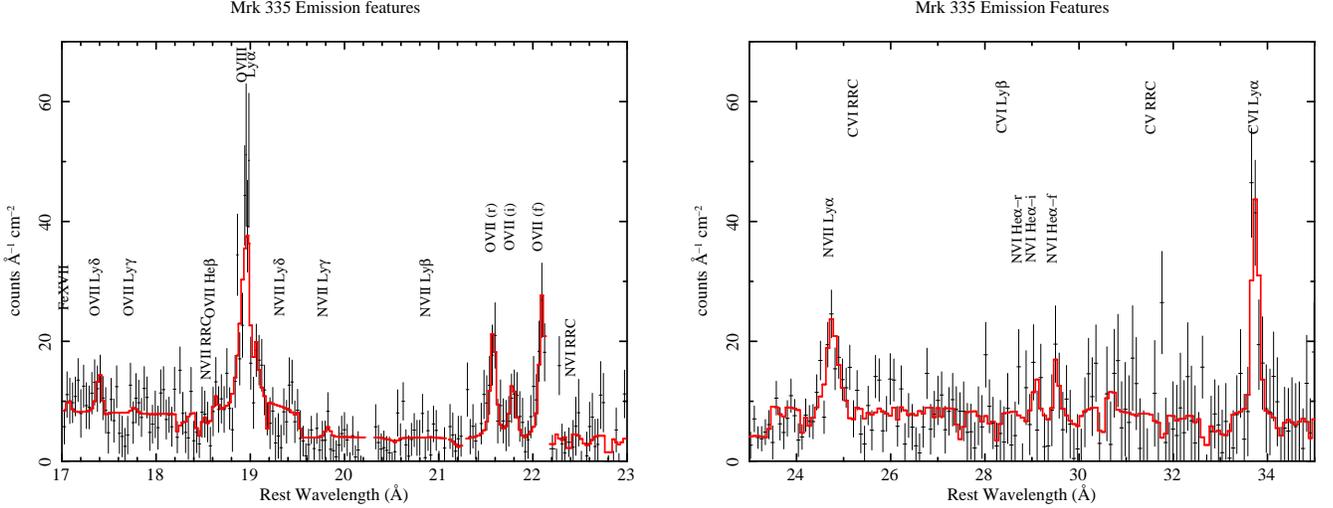

		\includegraphics[width=0.8\columnwidth,angle=-90]{fig4.ps}
		\includegraphics[width=0.8\columnwidth,angle=-90]{fig5.ps}
    \caption{Close up on the softest region of RGS spectrum. The labels mark only emission features, the warm absorber is included in the model but labels have been omitted to avoid overlapping with emission lines markers.}
    \label{fig:rgs_oxy}
\end{figure*}

To fit the soft X-ray continuum in the range 7-38~{\AA}  we start by including a power law with $\Gamma$=2.8  that is absorbed by a Galactic column density fixed to 3.6$\times$10$^{20}$ cm$^{-2}$ \citep{Kalberla05}. The Galactic absorption is modeled by the {\tt TBabs} component included in the suite of ISM absorption model developed by \cite{Wilms00}. 

\subsubsection{The emission features}
To this continuum, several narrow Gaussian lines are added to fit the emission lines. We initially included all the atomic transitions reported by \cite{Kinkha02} for the Seyfert 2 Galaxy NGC~1068. Many of them are detected in Mrk~335 spectrum and  the resulting list of emission lines with significance $\Delta$C$_{stat}$$>$1  (one free parameter) is reported in Table~\ref{tab_linelist}.  Line widths were initially set to 1~eV and line positions were fixed to their respective rest wavelength, therefore only the flux of the line is left as a free parameter. A close inspection of  the Ly$\alpha$ transitions from H-like ions lines in Fig.~\ref{fig:rgs_oxy} indicates that an underlying broader component is needed. Indeed, when the width of these lines  is left free, all of them are best described by a broad component with Full Width Hald Maximum (FWHM) of  around 2000~km~s$^{-1}$.  We also included Radiative Recombination Continua (RRC) features by adding 4 {\tt redge} components at the expected position of the CV, CVI, OVII and OVIII RRCs (see Table~\ref{tabRRC}).

 This model including only the emission component and the power law continuum yields C$_{stat}$/d.o.f.=3353/2998 and it is adopted in the following as a mere phenomenological description of the spectral features in emission.  A close up on the emission component is plotted in Fig.~\ref{fig:rgs_oxy}. 
   Further detailed analysis of the properties of the emission features is deferred to a forthcoming publication. 


\begin{table}[t]
\caption{\footnotesize{Emission lines detected in RGS spectrum.}}
\scriptsize
\begin{tabular}{c  c  c  c }
\hline\hline
 ID  & Rest             & Flux & $\Delta$Cstat \\ 
 -    &  Wavelength  & (10$^{-5}$ ph~cm$^{-2}$~s$^{-1}$) & - \\ 
   \hline 
   
 Mg XI He$\alpha$  {\it r}  &  9.228    & 0.247$^{+0.245}_{-0.224 }$  &   3.23  \\
 Ne X Ly$\alpha$  & 12.134    & 0.555$^{+0.298}_{-0.270 }$  &  13.12  \\
  NeIX He$\alpha$ {\it r} & 13.447    & 0.420$^{+0.228}_{-0.211 }$  &  11.37  \\
 NeIX {\it i}     & 13.553    & 0.120$^{+0.198}_{-0.120 }$  &   1.12  \\
  NeIX {\it f} & 13.698    & 0.391$^{+0.226}_{-0.210 }$  &  10.37  \\
  Fe XVII  & 15.015    & 0.301$^{+0.213}_{-0.193 }$  &   6.60  \\
  Fe XVII  & 17.073    & 0.225$^{+0.242}_{-0.224 }$  &   2.73  \\
  OVII He$\delta$ & 17.396    & 0.462$^{+0.275}_{-0.252 }$  &   9.57  \\
  OVII He$\gamma$   & 17.768    & 0.207$^{+0.259}_{-0.207 }$  &   1.93  \\
 OVII He$\beta$   & 18.627    & 0.215$^{+0.266}_{-0.215 }$  &   1.97  \\
   OVIII Ly$\alpha$  & 18.969    & 3.467$^{+0.536}_{-0.515 }$  & 146.87  \\
  NVII Ly$\gamma$ & 19.826    & 0.346$^{+0.418}_{-0.346 }$  &   2.37  \\
 OVII He$\alpha$ {\it r}   & 21.580    & 2.325$^{+0.833}_{-0.762 }$  &  28.97  \\
 OVII {\it i}    & 21.790    & 1.250$^{+0.693}_{-0.619 }$  &  12.04  \\
 OVII {\it f}      & 22.101    & 3.227$^{+0.979}_{-0.873 }$  &  50.15  \\
NVII Ly$\alpha$    & 24.781    & 2.270$^{+0.905}_{-0.992 }$  &  17.25  \\
 NVI He$\alpha$ {\it i}  & 29.083    & 0.609$^{+0.904}_{-0.609 }$  &   1.35  \\
   NVI {\it f}  & 29.534    & 1.163$^{+0.945}_{-0.840 }$  &   4.93  \\
  CVI Ly$\alpha$  & 33.736    & 7.314$^{+2.039}_{-1.894 }$  &  45.85  \\
\hline\hline
\end{tabular}
\label{tab_linelist}
\end{table}
 
 \begin{table}[h]
\begin{center}
\caption{\footnotesize{RRC in the RGS spectrum}}
\scriptsize
\begin{tabular}{c  c  c  }
\hline\hline
 ID   & Flux & $\Delta$Cstat \\ 
 -     & (ph~cm$^{-2}$~s$^{-1}$) & - \\ 
   \hline 
   
   CV   &       $<$ 3.21$\times$10$^{-5}$   &  1.4 \\
   
   CVI     &    $<$ 1.48$\times$10$^{-5}$    & 4.5  \\

 OVII   &     9.35$^{+3.58}_{-3.37}$$\times$10$^{-6}$   &  22 \\ 

 OVIII    &   3.44$^{+2.30}_{-2.14}$$\times$10$^{-6}$     &  9 \\ 
   
 \hline\hline
\end{tabular}
\end{center}
\label{tabRRC}
\end{table}

\subsubsection{The warm absorber}

The effect of line-of-sight absorption was then modeled by employing the suite of photoionization models {\tt PHASE} (Krongold et al. 2003) and after constructing the spectral energy distribution of the source that is necessary to calculate the ionization balance assumed to compute the absorption spectrum.  From  \cite{Longinotti13} we are well aware of the presence of a complex multi-component warm absorber in this source.
In order to allow a straightforward comparison of the present warm absorber properties with past epochs, we adopted the same spectral energy distribution described by  \cite{Longinotti13}. The UV-X-ray SED was built assuming the simultaneous fluxes and spectral shape from the Optical Monitor photometry and the EPIC-pn spectrum of 2009  shown in Fig.~4 of  \cite{Longinotti13}. This choice is also supported by the long term behavior of Mrk~335 reported by \cite{Gallo2018}, which is summarized in Section~\ref{sec:intro}.  

  The absorber in {\tt PHASE} is described by the following parameters: the gas ionization state defined as  U = $\frac{Q}{(4 \pi R^2 c n_e)}$  (with Q as the ionizing luminosity, n$_e$ as the gas electron density and R as the distance of the outflowing gas from the X-ray source),  the column density N$_H$, the turbulent velocity v$_{broad}$ and the outflow velocity. 
Indeed, the addition of an ionized absorber with initial best fit parameter log~U$\sim$0.8 and logN$_H$$\sim$21.8 produces an improvement of $\Delta$C$_{stat}$=~66 when compared to the model including the power law continuum and the emission component. 
 \begin{figure*}[t]
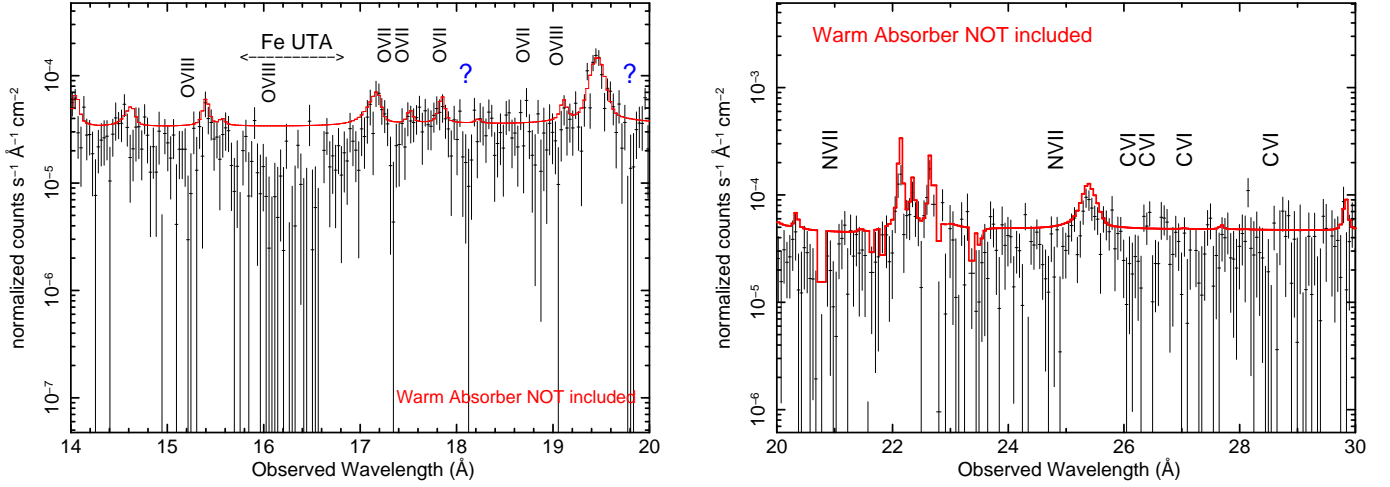

	\includegraphics[width=0.75\columnwidth,angle=-90]{fig6.ps}
	\includegraphics[width=0.75\columnwidth,angle=-90]{fig7.ps}
    \caption{Region of the RGS spectrum mostly affected by the warm absorber:  the spectrum is fitted only by the continuum plus the emission  component and labels mark the most prominent absorption features. The corresponding emission lines are displayed in Fig.~\ref{fig:rgs_oxy}.  Question marks highlight the positions of unidentified absorption lines. Data are binned for plotting purpose. }
    \label{fig:rgs_abs}
\end{figure*}

The detailed warm absorber parameters and errors are reported in Table~\ref{tab1}. 
The outflow velocity of this  absorber is of the order of 5-6$\times$10$^3$ km~s$^{-1}$, consistent with the velocity measured in the 2009 spectra by  \cite{Longinotti13}.  The velocity broadening of the lines in the absorber was initially set to 100~km~s$^{-1}$ and then left free to vary.  A  moderately tight constraint on this parameter could be found: v$_{broad}$~$\le$170~km~s$^{-1}$, therefore in the rest of the analysis this parameter is kept fixed to 100~km~s$^{-1}$. The slope of the underlying power law continuum is 2.72$\pm$0.18. 
Spectra plotted in Figure~\ref{fig:rgs_spec} and \ref{fig:rgs_oxy} include the effect of this layer of absorption. The final fit statistic for the model including the power law continuum, the  ionized absorber and the emission component is C$_{stat}$/d.o.f.=3287/2992.

Fig.~\ref{fig:rgs_abs} displays the most intense absorption features that are driving the warm absorber. The strongest absorption feature (left panel)  is due to a blend of several lines resulting from M-shell transitions in mildly ionized FeI-XVI, the so-called Fe UTA \cite[Unresolved Transition Array][]{Netzer04}. Further absorption is imprinted by transitions of CVI, NVII, OVII, OVIII. The question mark labels  plotted in the left panel of Fig.~\ref{fig:rgs_abs} mark the position of two unidentified absorption lines that could not be fitted self-consistently with another absorption component despite several attempts of finding a coherent model for these features. They will not be discussed in the remainder of this paper. 
 
An exhaustive comparison of the (several) multi-epoch X-ray data sets of Mrk~335 is out of the present scope and it will be the subject of a forthcoming publication. 
However, in Section~\ref{sec:discussion} we will review the properties of the present warm absorber (year 2015) compared to the findings reported by  \cite{Longinotti13} on the absorber emerged in Mrk~335 in the year 2009. 
  \begin{table}[h]
\begin{center}
\caption{\footnotesize{Best fit parameters of the RGS warm absorber.}}
\footnotesize
\begin{tabular}{c  c  c  c  }
\hline\hline
 Log U$^{(a)}$       &  Log N$_H$     & v$_{out}$   &   v$_{broad}$  \\
   -  &  (cm$^{-2}$) &        (km s$^{-1}$) &  (km s$^{-1}$)   \\
    \hline
           
 0.85$^{+0.09}_{-0.14}$  & 21.82$^{+0.20}_{-0.19}$ &   5700$^{+800}_{-400}$  & 100 (f)   \\ 

\hline\hline
\end{tabular} 

\label{tab1}
\end{center}
\end{table}



\subsection{The broadband  X-ray spectrum}
\label{sec:broadband}

The bandpass of the RGS instrument is limited to below 2.5~keV, therefore to achieve a full understanding  of the entire spectrum we applied the RGS best fit model  to the EPIC-pn data. This data is shown in Fig.\ref{fig:pn_spec}.
The RGS model offers a very detailed description of the ionized gas that is responsible for  emission and absorption in the soft X-ray band. Nonetheless the extension 
of  the bandwidth up to 10~keV immediately reveals the effect of unseen spectral components that are missing in this initial model due to bandpass limitation.

We start by adding to the RGS model a Gaussian emission line to accommodate strong residuals corresponding to a Fe~I~K$\alpha$ line that is highly prominent in this source (see details in Section~\ref{sub:Fekband}). We then added a blackbody component  ({\tt bbody}) to fit -at least phenomenologically- the strong soft excess that has always been present in X-ray data of Mrk~335 \cite[e.g.][]{Bianchi01,Grupe08,Gallo2015,Chainakun15}. This addition has the effect of flattening the underlying power law that is now extended to fit the hard X-ray band and that presents a photon index of $\Gamma$$\sim$0.9. The parameters of the blackbody component are in the range of the standard values for Seyfert~1 sources: the temperature is kT=0.11$\pm$0.02~keV. We note that the use of the more realistic  Comptonized blackbody model ({\tt comptt}) does not produce significant changes in the spectral fit, therefore we kept the more basic parametrization with {\tt bbody} to account for the soft excess. 
\begin{figure}[h]
\hspace{-0.8cm}	
\includegraphics[width=0.8\columnwidth,angle=-90]{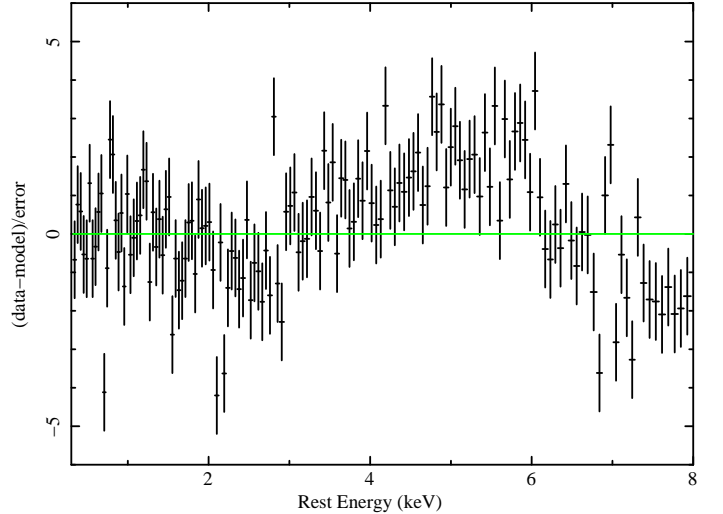}
    \caption{Residuals of the pn spectrum obtained before including the  partial covering and the reflection component in the continuum model (see \ref{sec:broadband}.}
    \label{fig:pn_res}
\end{figure}
Finally, guided by residuals around 1.7-2~keV, we added 4 Gaussian emission lines at the positions of 6.182, 6.740, 7.130, 8.421~\AA~to accommodate the transitions of SiXIV~Ly$\alpha$, SiXIII~He$\alpha$~{\it f}, Si~I~K$\alpha$ and MgXII~Ly$\alpha$.  These emission lines cannot be detected in the grating spectrum as they fall out of the bandpass. Nonetheless, considering the realm of features in Tables~\ref{tab_linelist} and \ref{tabRRC} along with examples of other AGNs where the emission spectrum could be measured in a wider band \cite[e.g. NGC~4151][]{Ogle00}, the presence of emission from heavier elements in the pn data of Mrk~335 is highly likely and indeed, their inclusion significantly improves the residuals in this spectral region. 

Even with these modifications, both the fit statistics of $\chi^2$/d.o.f.=416/141 and the hard X-ray curvature in the residuals (see Fig.\ref{fig:pn_res}) suggest the presence of  additional continuum component(s). 

\subsubsection{Partial covering absorber and Reflection component}
\label{sec:pcov_ref}
\begin{figure}[t]
\hspace{-0.8cm}
\includegraphics[width=1.2\columnwidth]{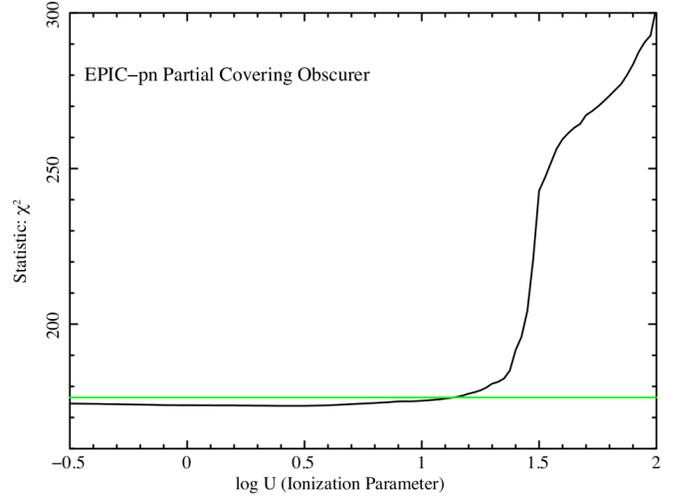}
	 \caption{Ionization parameter of the EPIC-pn partial covering absorber plotted against fit statistics of the best fit model ($\chi^2$/d.o.f.=174/138) that includes the reflection component (last paragraph of Section\ref{sec:broadband}. }
    \label{fig:pcov_u}
\end{figure}

To mimic the effect of mildly ionized gas  partially covering the primary X-ray continuum we applied to the power law  an additional layer of absorption parametrized by a second {\tt PHASE} component with an initial low ionization parameter 
and a variable covering factor.  
This partial covering component significantly improves the spectral fit ($\Delta\chi^2$=220 for 3 d.o.f.) and the intrinsic power law gets to a steeper  photon index more typical of  Mrk~335, $\Gamma$=1.65$\pm$0.11.  The column density of this gas is found to be quite high, log(N$_H$)=22.99$\pm$0.06, and the covering factor is 0.79$^{+0.02}_{-0.05}$. The ionization parameter could not be constrained precisely (see Fig. \ref{fig:pcov_u}) but the  upper limit of logU  points to a degree of ionization lower than logU$\sim$1.35.   Likewise, the velocity of this absorber could not be measured  due to the limited resolution of the pn CCD, therefore we kept it fixed to the same value of the RGS  warm absorber ($\sim$5600~km~s$^{-1}$). We note that testing alternative velocities (e.g. v$_{out}$=2000 or 800 km~s$^{-1}$) does not provide any relevant change in the spectral fit.

During the fitting process, some parameters of the  ionized absorber detected in the RGS have been frozen: the ionization parameter, outflow velocity and velocity broadening are frozen to the best fit values reported in Table~\ref{tab1}, while the column density is left free.  This is justified by considering that the coarser resolution of the CCD cannot improve the parameters already well constrained by grating spectroscopy. 

\begin{figure}[t]
\hspace{-0.8cm}
\includegraphics[width=0.8\columnwidth,angle=-90]{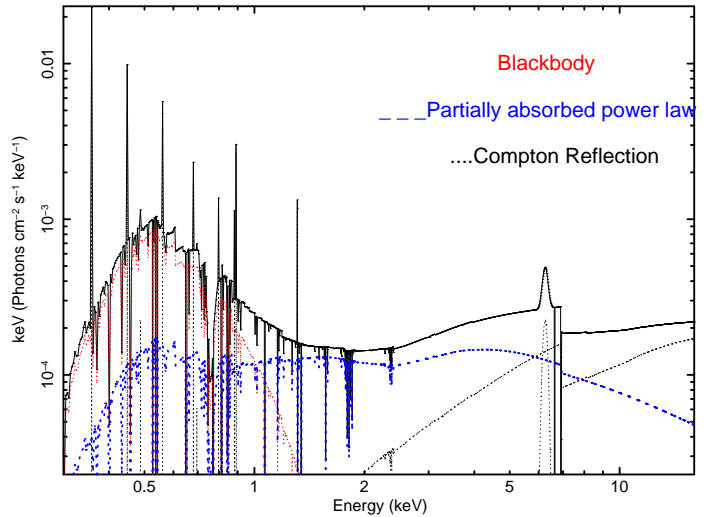}
\caption{Broadband model applied to the EPIC-pn spectrum. Solid black: combination of all spectral components. Red: Blackbody + warm absorber detected in RGS. Blue: power law absorbed by mildly ionized gas covering $\sim$80\% of the X-ray source and by warm absorber. Dotted black: Compton Reflection. Soft X-ray emission features are modeled by Gaussian lines. For simplicity, the Fe K absorption is also modeled by two Gaussian lines. }
    \label{fig:broadband}
\end{figure}

As a conclusive and necessary step of the analysis of the broadband continuum, we also considered the presence of a Compton reflection component. A detailed analysis of the reflection spectrum and the property of the inner accretion disc of Mrk~335 is out of the scope of this paper and it is presented elsewhere  \citep{Gallo2019}. However, the presence of inner relativistic reflection in Mrk~335 was intensively studied in recent years \citep{Kara2013,Parker2014,Gallo2015} and eventually confirmed as one of the dominant spectral component of this AGN. Therefore we included a basic parameterization of the reflection spectrum by removing the partial covering and by adding a {\tt pexrav} 
component to the broadband model.  This test yields a reduced $\chi^2$ of 3.38 and a much flatter power law  ($\Gamma$$\sim$1.26),  indicating  that partial obscuration  is still required by the data.  Once the partial covering  is included back into the model with the reflection component, the slope of the continuum goes to $\Gamma$=2.14$^{+0.10}_{-0.13}$ and the fit statistics improves to $\chi^2$/d.o.f.=174/138. The broadband model is plotted in Figure~\ref{fig:broadband}.  We remark that this parametrization serves merely to test the statistical requirement of the partially covering gas, therefore a detailed spectral fitting is not envisaged herein and standard reflection parameters are adopted: the cut-off energy is 500 keV, solar abundances are chosen for the elements heavier than He and for Fe, and the inclination angle of the disk is fixed to 30 degrees. The only fitting parameter left free is the reflection fraction that, not surprisingly,  pegs to its maximum value (R=10) indicating a dominant contribution from the inner accretion disk. These values are broadly  consistent with what reported by \cite{Parker2014}, who, in their relativistic treatment, had found high reflection fractions in NuSTAR data of Mrk 335 at very low flux state. This behavior is also reported by the recent publication by  \cite{Gallo2019}, to which the reader is deferred for a more detailed analysis of the reflection properties of the source. Finally, we note that our coarse parametrization does not exclude the likely contribution of a more distant reflector from the outer part of the disk or the molecular torus of the AGN, as indicated by the narrow Fe~K$\alpha$ line reported in next section.

\subsubsection{The Iron line band}
\label{sub:Fekband}
We now take a closer look at the Fe~K band. Owing to the spectral complexity of the soft X-ray band and with the aim to speed up the fitting procedure, the following analysis was carried out on the data within the range 3-8 keV. This is justified by considering that the opacity of the soft X-ray warm absorber has no effect above 3 keV and that the bulk of  strong emission lines are emitted below this  energy threshold. The continuum model from the previous section constituted by a partially covered power law plus Compton reflection has been applied to this restricted band (the blackbody component was dropped since it has no effect in this band). The choice of not extending the bandwidth to the nominal 10~keV is due to the rising of instrumental background above 8~keV that introduces significant uncertainty in the spectral features apparently present above this threshold. 

\begin{figure}[t]
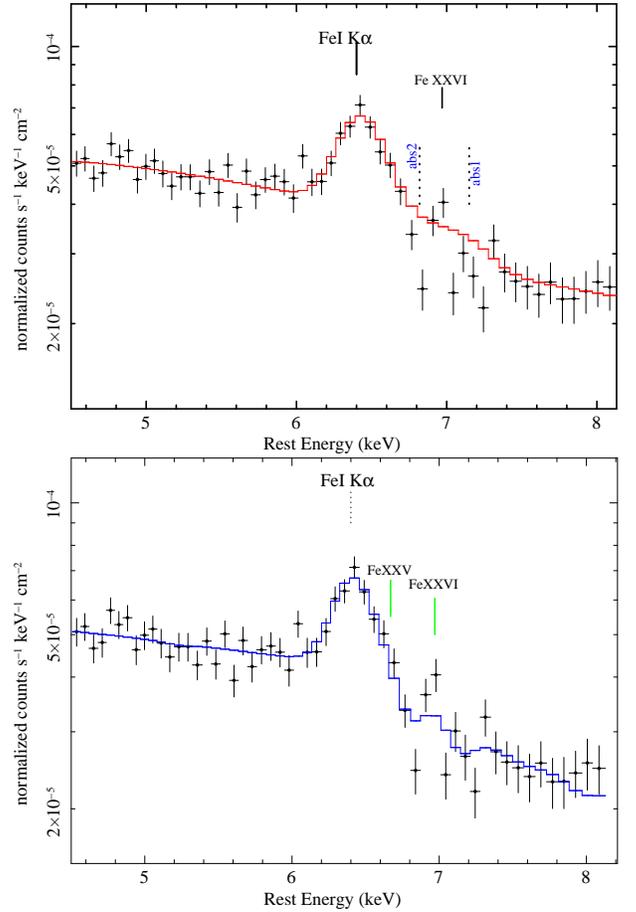

\centering
\resizebox{1.0\hsize}{!}{\includegraphics[angle=270]{fig11.ps}}
\resizebox{1.0\hsize}{!}{\includegraphics[angle=270]{fig12.ps}} 
	 \caption{Close up of the Fe K band in the EPIC pn spectrum. Top:  the underlying continuum is modeled as a partially covered power law and only one Gaussian emission line is included to model the Fe~I K$\alpha$ line. The laboratory position of the FeXXVI transition is marked by the solid line in black color (see  \ref{sub:Fekband}) Dotted vertical lines mark the position of the two absorption lines (blue vertical labels). Bottom: the spectrum is modeled as above but now including the {\tt PHASE} component of the high ionization absorber with outflow velocity of 5,200$^{+700}_{-200}$~km~s$^{-1}$ described at the end of \ref{sub:Fekband}. Green labels mark the laboratory positions of the transitions of highly ionized Fe.} 
    \label{fig:fek}
\end{figure}

 The presence of the K$\alpha$ emission line from neutral Iron is very evident in the spectrum and it has been fitted with a Gaussian profile with peak energy E=6.41$^{+0.03}_{-0.02}$~keV and width $\sigma$=0.12$\pm$0.03~keV.  The intensity of the Fe~K$\alpha$ line parametrized with this Gaussian profile and expressed as its Equivalent Width is EW=300$\pm$45~eV. 
 The continuum model after the inclusion of the Fe~K$\alpha$ line yields a fit statistics of $\chi^2$/d.o.f.=97/74. The spectrum fitted by this model is plotted in Fig.~\ref{fig:fek}. 
  The Fe line parameters are compatible with emission in the molecular torus via Compton reflection, as proposed by \cite{O'Neill07} for the high flux state of Mrk 335. We note that the contribution of a distant reflector was not directly tested via spectral fitting  in the present data but it is discussed in  \cite{Gallo2019}.
 
 Additional residuals on the blue side of the Fe~K$\alpha$ line suggest us to explore the presence of emission from highly ionized Iron, which was already revealed when the source was observed in high flux state with higher photon statistics \citep{O'Neill07}.
 We added a narrow ($\sigma$=1 eV) Gaussian line in emission and measured its position at E=6.96$^{+0.05}_{-0.16}$~keV, but only an upper limit of EW$\le$10~eV could be measured, therefore this line is no longer included in the following tests. 

We then proceed to examine the residuals in absorption that are still present in the spectrum. Indeed the addition of a narrow ($\sigma$=1~eV) Gaussian line with negative intensity at a redshift-corrected position of E=7.15$\pm$0.09~keV (``abs1" in Fig.~\ref{fig:fek})  improves the fit statistics by $\Delta\chi^2$=8 (for 2 d.o.f.) and its intensity  is measured to EW=57$\pm$30~eV. 
A second absorption line is found at the position of E=6.82$\pm$0.05~keV (``abs2" in Fig.~\ref{fig:fek}) with an intensity of EW=68$\pm$25~eV and statistical improvement of $\Delta\chi^2$=10 for 2 d.o.f.

These absorption features suggest the presence of an ionized blue-shifted absorber  that could constitute  a high ionization layer  of the outflowing system detected in the  soft X-ray. The closest transitions that could originate the FeK absorption feature at 7.15~keV are Fe~XXVI (E$_{lab}$=6.97~keV) and Fe XXV (E$_{lab}$=6.67~keV).  The corresponding outflow velocity would be respectively $\sim$~7,500 and 20,000~km~s$^{-1}$. 
With regard to the second absorption line at 6.82~keV, if interpreted as blue-shifted FeXXV, the outflow velocity would be around $\sim$7000~km~s$^{-1}$. Considering these numbers and the large uncertainties in the position of both absorption lines measured in the pn CCD data, the most viable interpretation is that both absorption lines in the Fe K band correspond to He and H-like Iron originated in a gas outflowing at a velocity of $\sim$ 7000~km~s$^{-1}$,  in reasonable agreement  with the velocity pattern  (5700~km~s$^{-1}$) of the warm absorber detected in the grating spectrum.

As a final step, we have replaced the two absorption lines in the spectral model with a {\tt PHASE} component that can self-consistently fit both features. The improvement in the fit statistics corresponds to $\Delta\chi^2$=19 for 3 d.o.f., and the parameters of the photoionized wind are logU=3.13$^{+0.09}_{-0.59}$, logN$_H$$\ge$23.07, with an outflow velocity of 5,200$^{+700}_{-200}$~km~s$^{-1}$. This velocity, which is now measured self-consistently, shows a much finer agreement with the value derived from the grating spectrum for the low ionization component of the absorber.

\subsubsection{On the consistency of the broadband model with the grating spectrum}
Due to bandwidth restriction, neither of the two EPIC-pn  absorbers (the highly ionized and the partial covering) seem to imprint obvious features in the RGS data. 

With regard to the partial covering, the bulk of absorption comes from  continuum opacity and its effect starts to be visible in the spectrum above 3 keV, therefore virtually impossible to be detected in RGS. 
The inclusion of a  {\tt PHASE} component in the RGS best fit model with parameters fixed to the EPIC-pn values is formally consistent with the data, although no improvement in the fit statistics is found.  When the column density and the velocity are kept frozen, logU is found around 1.04 with an extremely low covering factor (C$_f$=0.1), that would not allow any individual absorption line to be strong enough to be detected in RGS.

We now explore the possible presence of the EPIC-pn highly ionized absorber in RGS data. The inclusion of this component into the RGS best fit model  yields a modest  improvement of $\Delta\chi^2$=5 for 3 free parameters: logU=2.61$^{+0.75}_{-0.33}$, logN$_H$=23.4$^{+0.39}_{-0.99}$ and v$_{out}$=6500$^{+400}_{-700}$~km~s$^{-1}$. These parameters are broadly consistent with the EPIC-pn values reported at the end of the previous section. 

We fully acknowledge that  the presence of the high ionization absorber is not statistically robust in any of the spectral data.  Nonetheless, after running these checks,  we conclude that the simultaneous (albeit moderate) significance in CCD and grating spectra of an outflow with velocity consistent with the well characterized system of winds observed in RGS,  concurs to indicate that a high ionization component of the wind is present in the low flux state of Mrk~335.

\section{Hubble Space Telescope Ultraviolet Spectra}

\subsection{Observations and Data Reduction}

The triggered {\it XMM-Newton} observations of Mrk 335 were coordinated with
two {\it HST} observations. The first of these, on 2016-01-04, followed the
{\it XMM-Newton} observation by five days. This visit used all COS FUV
gratings (G130M, G160M, and G140L) to cover the full wavelength range
from 912 \AA\ to 2000 \AA, specifically including the region surrounding
Ly$\beta$ and \ion{O}{6}.
The second visit, another three days later, on 2016-01-07, supplemented the
G140L exposures to obtain better S/N in the Ly$\beta$ and \ion{O}{6} region.
Table \ref{ObsTbl} gives the observational details of the individual spectra.
\citet{Green12} describe the key characteristics of the design and performance
of the COS instrument on the Hubble Space Telescope (HST).
The G130M and G160M gratings have a resolving power of $R \sim 15\,000$ over
the wavelength range of 1135--1800 \AA.
The G140L grating has resolving power $R \sim 2\,000$ covering
912 \AA\ to 2000 \AA\ with the 1280 central wavelength setting, but
with a gap between detector segments from 1190--1265 \AA.
We chose two central wavelength settings for G130M and G160M to bridge the gap
between the FUV detector segments. These settings were also chosen to avoid
placing the gap on spectral features of interest in Mrk 335.
In addition to multiple central wavelength settings, we also used multiple
focal-plane positions to avoid flat-field features and other detector
artifacts.

\begin{table}
  \centering
        \caption{COS Observations of Mrk 335}
        \label{ObsTbl}
        \scriptsize
\begin{tabular}{l c c c c}
\hline\hline
Data Set  & Grating/Tilt  & Date & Start Time & Exposure \\
              &               &      &    (GMT)   & (s)\\
\hline
lckg01010 & G130M/1291 & 2016-01-04 &  03:06:56 &  $\phantom{0}450$ \\
lckg01020 & G130M/1318 & 2016-01-04 &  03:17:49 &  $\phantom{0}440$ \\
lckg01030 & G130M/1327 & 2016-01-04 &  03:28:26 &  $\phantom{0}880$ \\
lckg01040 & G160M/1611 & 2016-01-04 &  04:35:27 &  $\phantom{0}800$ \\
lckg01050 & G160M/1623 & 2016-01-04 &  05:04:41 &  $\phantom{0}800$ \\
lckg01060 & G140L/1280 & 2016-01-04 &  06:10:51 &  2320 \\
lckg01070 & G140L/1280 & 2016-01-07 &  01:07:30 &  1740 \\
lckg02010 & G140L/1280 & 2016-01-07 &  02:43:57 &  5161 \\
\hline
\end{tabular}

\end{table}

\begin{figure*}
  \centering
   \includegraphics[width=17cm, angle=-90,, scale=0.45, trim=0 0 230 0]{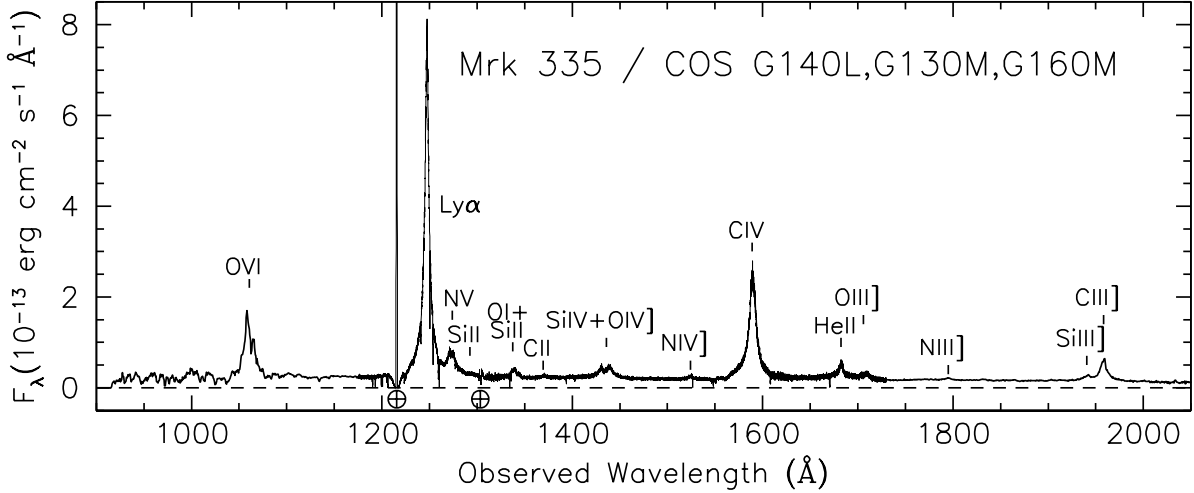}
  \caption{Calibrated COS spectra of Mrk 335 from HST observations on
2016-01-04 and 2016-01-07.
Wavelengths shortward of 1180 \AA\ and longward of 1780 \AA\ are from the
merged G140L observations. Between 1180 \AA\ and 1423 \AA, the data are from
G130M. Between 1423 \AA\ and 1780 \AA, the data are from G160M.
We mark the prominent emission lines with their identifications.
Earth symbols mark geocoronal emission from
Ly$\alpha$ and \ion{O}{1}$\lambda$1302.
}\label{fig_cosfull}
\end{figure*}

We processed the observations using v3.1 of CALCOS, the COS calibration
pipeline, supplemented by custom flat-field files developed for lifetime
position 3.
The wavelength zero points of all spectra were adjusted after processing
by measuring the wavelengths of strong interstellar features and aligning
them to the line-of-sight velocity for \ion{H}{1},
$V_{LSR} = -11~\rm km~s^{-1}$ \citep{Murphy96}.
Improvements in the COS wavelength calibration now give relative
uncertainties of $\sim 5~\kmps$.
Comparing the eight exposures in Table \ref{ObsTbl}, fluxes in the wavelength
regions each has in common agree to better than 2\%, and within the
statistical errors of each exposure.
Therefore, we combine all exposures for each grating to make three separate
spectra for G130M, G160M, and G140L, and we join G130M and G160M at
1423 \AA\ to make a single high resolution spectrum covering 1135--1800 \AA.
Figure \ref{fig_cosfull} shows the full merged COS spectrum from the Lyman
limit to 2000 \AA.

\subsection{Measuring the Broad UV Absorption Lines}
\label{sec:UV_abslines}

Unlike our discovery of broad \ion{C}{4} absorption in the 2009 and 2010
COS spectra of Mrk 335 \citep{Longinotti13}, in our new observation
we know where to look to see if the absorption features have reappeared
in our triggered observations of Mrk 335 in an obscured state.
Upon close inspection, broad absorption in \ion{C}{4} and Ly$\alpha$ were
immediately apparent in our 2016 observations.
To measure the properties of these absorption features,
we first developed a total emission model for Mrk 335 that includes both
the continuum and the emission lines.
As in \cite{Longinotti13}, we use a power law for the continuum,
$\rm F_\lambda = F_{1000}~ (\lambda / 1000 \AA)^{-\alpha}$
reddened by E(B-V) = 0.030 \citep{Schlafly11} using the \cite{CCM} extinction
law with $\rm R_V = 3.1$.
Our model includes more components than in \cite{Longinotti13}, as we include
components for weaker emission lines as well as  \ion{Si}{4}, and the
\ion{C}{3} $\lambda977$, \ion{N}{3} $\lambda991$,
\ion{Si}{3}] $\lambda1892$, and \ion{C}{3}] $\lambda1909$ lines present in the
the G140L spectrum.

We fit our model in three separate pieces due to the complexity of the
spectrum and the greatly differing resolutions of the G130M+G160M and the G140L
gratings. We first fit G130M+G160M, covering 1140--1800 \AA\ as described
below, then separately fit the
\ion{Si}{3}] $\lambda1892$, and \ion{C}{3}] $\lambda1909$
region at the red end of G140L, and the Ly$\beta$+\ion{O}{6} region on
the blue end.
For the G140L fits we constrain the power-law continuum to have the same
spectral index as in the fit to the G130M+G160M spectrum, but we allow its
normalization to adjust to the slight ($\sim2$\%) differences relative to
G130M and G160M.

In \cite{Longinotti13}, the restricted wavelength range of the G130M and
G160M spectra did not include a substantial contribution from the forest of
\ion{Fe}{2} emission lines that blend into a pseudo-continuum starting at
roughly 1500 \AA.
This pseudo-continuum is brighter in our current G140L spectrum, but there are
no definitive spectral features that enable us to constrain its strength.
These lie mostly at longer wavelengths, peaking at $\sim$2500 \AA.
We therefore used the FOS spectrum described in
\cite{Longinotti13} to constrain its normalization, and included it as a
fixed element of our model, using a scaled version of the \cite{Wills85}
model convolved with a Gaussian full-width at half maximum (FWHM) of
2800 $\rm km~s^{-1}$, approximately the FWHM of the \ion{C}{4} emission line.
At 1800 \AA, its flux is $3 \times 10^{-16} \rm ~\ergcmsA$, which is only
2\% of the modeled continuum in our spectrum.

We use multiple Gaussian components for each line, choosing an
appropriate number to obtain the best fit for each line.
These components have only rough physical meanings. We presume that the
narrow components may be more reflective of the narrow-line region (NLR) in
Mrk 335, but we ascribe no significance to the broader components we have used
to decompose each profile. We are merely interested in a good characterization
of the total emission, which will enable us to normalize the spectrum,
measure the strengths of the absorption features, and measure the total
flux of the emission components above the continuum.
The strongest emission lines in Mrk 335 require at least three, and sometimes
four components. The narrowest portions of the profiles are clearly separated in
\ion{O}{6}, \ion{N}{5}, \ion{Si}{4}, and \ion{C}{4}, so we
include a separate component for each line of the doublet.
We link the wavelengths of the doublet lines at the ratio of their
vacuum wavelengths, and we fix the ratio of their fluxes at a ratio of 2:1,
assuming they are optically thin.
Ly$\alpha$, \ion{N}{5}, \ion{Si}{4}, \ion{C}{4}, and \ion{He}{2}
all require a very broad Gaussian component with full-width at half maximum
(FWHM)$\sim$10\,000 $\rm km~s^{-1}$.
We do not model this separately for the doublets in \ion{N}{5}, \ion{Si}{4},
and \ion{C}{4}.
\begin{longtable*}{c c c c c}
        \caption{{\footnotesize Emission Features in the 2016 COS Spectra of Mrk 335. Parameters with error bars of zero were tied to other parameters in the fit (see \ref{sec:UV_abslines}).}}
        \label{COS_elines}  \\
\hline
\hline
Feature & $\rm \lambda_0$ & Flux & $\rm v_{sys}$ & FWHM  \\
  & ($\rm \AA$)  & ($\rm 10^{-14}~erg~cm^{-2}~s^{-1}~\AA{-1}$) & ($\rm km~s^{-1}$) & ($\rm km~s^{-1}$) \\
\hline
\hline

\endfirsthead
\multicolumn{5}{c}%
{ \tablename\ \thetable{} -- continued from previous page} \\
\hline\hline
Feature & $\rm \lambda_0$ & Flux & $\rm v_{sys}$ & FWHM \\
  & ($\rm \AA$)  & ($\rm 10^{-14}~erg~cm^{-2}~s^{-1}~\AA{-1}$) & ($\rm km~s^{-1}$) & ($\rm km~s^{-1}$) \\
\hline
\hline
\endhead

\hline \multicolumn{5}{c}{{ -- Continued on next page}} \\ \hline
\endfoot

\hline \hline
\endlastfoot

\ion{C}{3} &  977.02 & $ 13.9\pm 5.9$ & $ -513 \pm  82$ & $ 3271 \pm  143$\\
\ion{N}{3} &  990.82 & $  7.5\pm 5.5$ & $ -197 \pm 101$ & $ 3000 \pm  170$\\
\ion{O}{6} & 1031.93 & $  3.9\pm  7.3$ & $  -114 \pm  57$ & $  700 \pm   90$\\
\ion{O}{6} & 1037.62 & $  4.8\pm  1.4$ & $  -113 \pm   0$ & $  700 \pm    0$\\
\ion{O}{6} & 1031.93 & $ 40.7\pm  2.1$ & $   -38 \pm   0$ & $ 1193 \pm  130$\\
\ion{O}{6} & 1037.62 & $ 34.2\pm  3.4$ & $   -37 \pm   0$ & $ 1193 \pm    0$\\
\ion{O}{6} & 1031.93 & $ 29.1\pm  0.1$ & $  -123 \pm   0$ & $ 6210 \pm  220$\\
\ion{O}{6} & 1037.62 & $ 29.1\pm  0.0$ & $  -125 \pm   0$ & $ 6210 \pm    0$\\
\ion{O}{6} & 1034.78 & $ 50.2\pm  0.0$ & $  -130 \pm   0$ & $ 8736 \pm    0$\\
\ion{S}{4} & 1072.97 & $  4.3\pm  1.3$ & $   234 \pm 120$ & $ 2900 \pm  180$\\
\ion{P}{5} & 1117.98 & $  1.7\pm  1.0$ & $     1 \pm 134$ & $ 4393 \pm    0$\\
\ion{P}{5} & 1128.01 & $  1.0\pm  1.0$ & $     1 \pm   0$ & $ 4393 \pm    0$\\
\ion{C}{3}* & 1176.01 & $  0.4\pm  1.1$ & $    -3 \pm  49$ & $ 1334 \pm   96$\\
\ion{C}{3}* & 1176.01 & $ 13.6\pm  2.2$ & $    -3 \pm  45$ & $ 4393 \pm  160$\\
Ly$\alpha$ & 1215.67 & $104.0\pm  2.0$ & $    92 \pm   5$ & $  687 \pm   10$\\
Ly$\alpha$ & 1215.67 & $219.0\pm  4.1$ & $    54 \pm   5$ & $ 1724 \pm   20$\\
Ly$\alpha$ & 1215.67 & $194.0\pm  4.2$ & $   112 \pm   5$ & $ 4892 \pm  118$\\
Ly$\alpha$ & 1215.67 & $ 78.3\pm  2.3$ & $    92 \pm  14$ & $10743 \pm  196$\\
\ion{N}{5} & 1238.82 & $  4.2\pm  1.1$ & $    25 \pm   5$ & $  702 \pm   11$\\
\ion{N}{5} & 1242.80 & $  4.2\pm  0.0$ & $    24 \pm   0$ & $  702 \pm    0$\\
\ion{N}{5} & 1238.82 & $ 18.6\pm  2.0$ & $    56 \pm   5$ & $ 2530 \pm   95$\\
\ion{N}{5} & 1242.80 & $ 18.6\pm  0.0$ & $    57 \pm   0$ & $ 2530 \pm    0$\\
\ion{N}{5} & 1240.89 & $ 53.0\pm  1.2$ & $    27 \pm  18$ & $ 8715 \pm  115$\\
\ion{Si}{2} & 1260.42 & $  3.0\pm  1.0$ & $   574 \pm  41$ & $ 1724 \pm    0$\\
\ion{O}{1}+\ion{Si}{2} & 1304.46 & $  1.4\pm  1.1$ & $  -420 \pm   5$ & $  800 \pm  133$\\
\ion{O}{1}+\ion{Si}{2} & 1304.46 & $ 14.5\pm  1.2$ & $   -21 \pm   5$ & $ 3000 \pm  113$\\
\ion{C}{2} & 1334.53 & $  1.3\pm  1.0$ & $   313 \pm  16$ & $  800 \pm    0$\\
\ion{C}{2} & 1334.53 & $  5.0\pm  1.0$ & $   109 \pm   8$ & $ 3000 \pm    0$\\
\ion{Si}{4} & 1393.76 & $  4.7\pm  1.1$ & $   110 \pm  13$ & $ 1021 \pm  112$\\
\ion{Si}{4} & 1402.77 & $  4.7\pm  0.0$ & $   110 \pm   0$ & $ 1021 \pm    0$\\
\ion{Si}{4} & 1393.76 & $  6.6\pm  1.1$ & $   -61 \pm  14$ & $ 3645 \pm  135$\\
\ion{Si}{4} & 1402.77 & $  6.6\pm  0.0$ & $   -63 \pm   0$ & $ 3645 \pm    0$\\
\ion{Si}{4} & 1398.19 & $ 24.3\pm  1.1$ & $ -1045 \pm  15$ & $10413 \pm  149$\\
\ion{O}{4}] & 1401.16 & $  1.9\pm  1.1$ & $   253 \pm  92$ & $ 1021 \pm    0$\\
\ion{O}{4}] & 1401.16 & $  9.1\pm  1.3$ & $   253 \pm   0$ & $ 3645 \pm    0$\\
\ion{N}{4}] & 1486.50 & $  2.1\pm  1.1$ & $    13 \pm  12$ & $ 1021 \pm    0$\\
\ion{N}{4}] & 1486.50 & $  3.7\pm  1.0$ & $   -84 \pm  17$ & $ 2590 \pm  154$\\
\ion{C}{4} & 1548.19 & $ 30.8\pm  2.0$ & $    39 \pm   5$ & $  811 \pm   22$\\
\ion{C}{4} & 1550.77 & $ 30.8\pm  0.0$ & $    38 \pm   0$ & $  811 \pm    0$\\
\ion{C}{4} & 1548.19 & $ 39.6\pm  3.0$ & $    56 \pm   5$ & $ 1904 \pm  112$\\
\ion{C}{4} & 1550.77 & $ 39.6\pm  0.0$ & $    55 \pm   0$ & $ 1904 \pm    0$\\
\ion{C}{4} & 1548.19 & $ 43.1\pm  3.0$ & $  -360 \pm   8$ & $ 4294 \pm  113$\\
\ion{C}{4} & 1550.77 & $ 43.1\pm  0.0$ & $  -360 \pm   0$ & $ 4294 \pm    0$\\
\ion{C}{4} & 1549.05 & $ 74.4\pm  3.1$ & $    56 \pm  14$ & $ 8736 \pm  119$\\
\ion{He}{2} & 1640.48 & $  8.9\pm  1.1$ & $     5 \pm   5$ & $  779 \pm   16$\\
\ion{He}{2} & 1640.48 & $  8.8\pm  1.1$ & $  -200 \pm   5$ & $ 2272 \pm  115$\\
\ion{He}{2} & 1640.48 & $ 57.1\pm  2.0$ & $ -1373 \pm  22$ & $11989 \pm  288$\\
\ion{O}{3}] & 1660.81 & $  3.0\pm  1.0$ & $    64 \pm   5$ & $ 1600 \pm   43$\\
\ion{O}{3}] & 1666.15 & $  8.0\pm  1.0$ & $    64 \pm   0$ & $ 1600 \pm    0$\\
\ion{N}{3}] & 1750.00 & $  1.1\pm  1.0$ & $    -1 \pm  68$ & $ 3267 \pm  324$\\
\ion{Si}{3}] & 1892.08 & $ 6.1\pm  0.9$ & $    67 \pm  70$ & $ 1169 \pm  161$\\
\ion{Si}{3}] & 1892.08 & $ 5.6\pm  2.0$ & $  -635 \pm 541$ & $ 2920 \pm  121$\\
\ion{C}{3}] & 1908.73 & $ 16.3\pm  3.4$ & $    12 \pm  19$ & $  850 \pm   92$\\
\ion{C}{3}] & 1908.73 & $ 31.6\pm  0.9$ & $   102 \pm  56$ & $ 2677 \pm  506$\\

\hline
\end{longtable*}

To optimize our fits, we used the
IRAF\footnote{IRAF (http://iraf.noao.edu/) is distributed by the
National Optical Astronomy Observatory, which is operated by the
Association of Universities for Research in Astronomy, Inc.,
under cooperative agreement with the National Science Foundation.}
task {\sc specfit} \citep{Kriss94}.
We first fit the merged G130M+G160M spectra.
Our best-fit model has a power-law normalization of
$\rm F_{1000} = 4.09 \times 10^{-14}~\ergcmsA$ with a spectral index of
$\alpha = 1.31$.
The fit is excellent, with $\chi^2=12049.76$ for 12285
points and 99 free parameters.
The best fit parameters are given in Table~\ref{COS_elines}.
Error bars are calculated from the error matrix of the fit, assuming
a 1$\sigma$ error corresponds to $\Delta \chi^2 = 1$ for a single
interesting parameter.
Parameters with error bars of zero were tied to other parameters in the fit,
e.g., the fluxes, wavelengths, and widths of doublet emission lines, and
the widths of some of the weakest emission components.

\begin{figure}[t]
\centering
\resizebox{1.0\hsize}{!}{\includegraphics[angle=270]{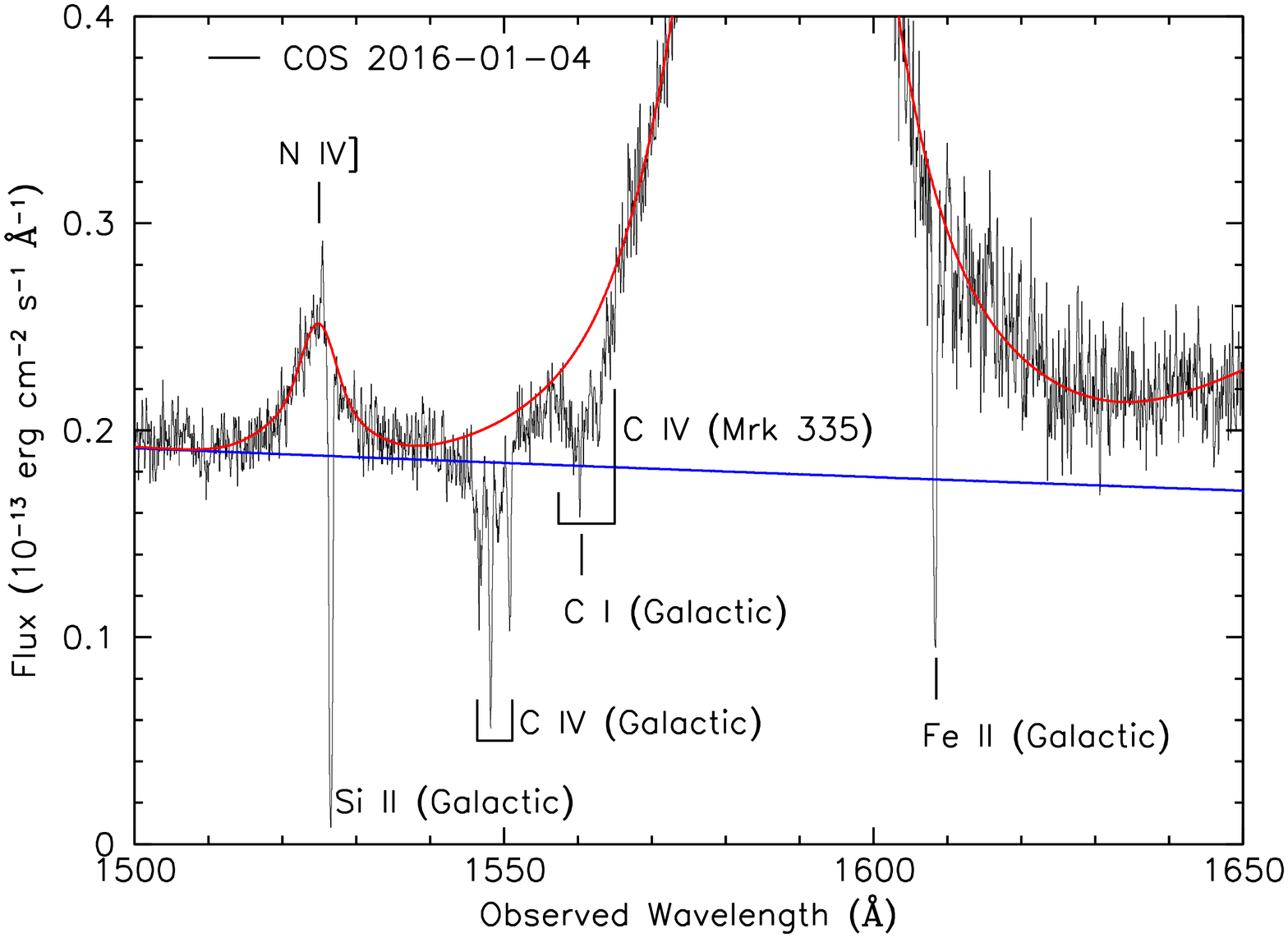}}
\resizebox{1.0\hsize}{!}{\includegraphics[angle=270]{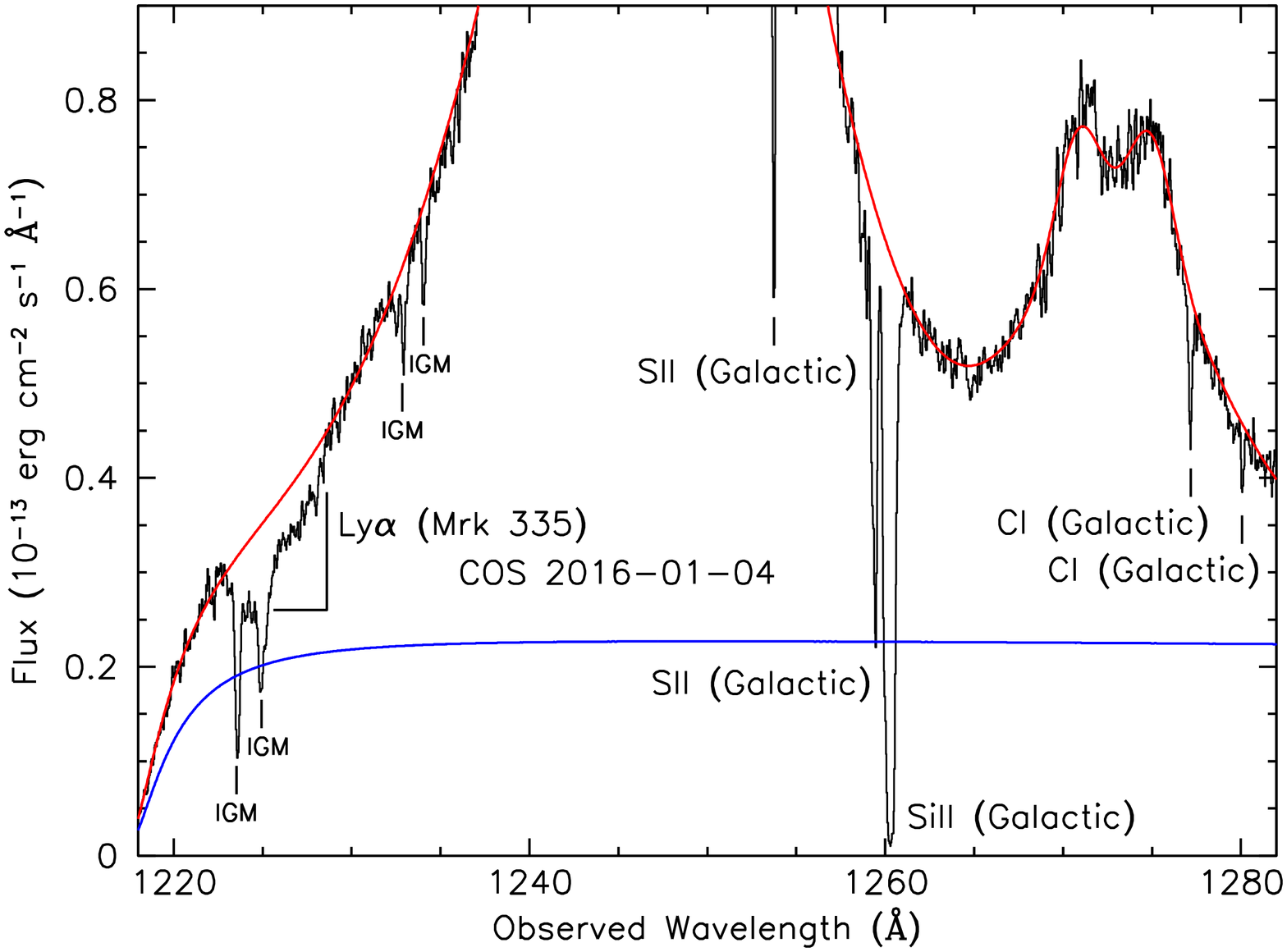}}
\caption{
{\it Top}: Best fit to the COS G160M 2016 spectrum of the \ion{C}{4} region
binned by 4 pixels (black histogram). {\it Bottom}: Best fit to the COS G130M 2016 spectrum of the Ly$\alpha$+\ion{N}{5} region
binned by 4 pixels (black histogram).
The peak of the \ion{C}{4}  (top)  and Ly$\alpha$  (bottom) emission lines  has been truncated to show more detail
at the flux levels of the blue-shifted broad absorption.
The solid red line tracing the data is the total emission model.
The continuum component is the solid blue line (including foreground damped Milky Way Ly$\alpha$
absorption).
Foreground interstellar absorption lines  and absorption due to the
intergalactic medium (IGM) are marked.
The intrinsic broad \ion{C}{4} (Ly$\alpha$) absorption is marked as the depression below
the emission model  (at 1227 \AA) surrounding the narrow Galactic \ion{C}{1} absorption line.
}
\label{fig_c4fit}
\end{figure}

Strong, broad absorption in Ly$\alpha$ and \ion{C}{4} is clearly present in
our new observations.
Figure \ref{fig_c4fit} shows our best fit to the \ion{C}{4} region.
Absorption is clearly present on the extreme blue wing of the \ion{C}{4}
emission profile (in the same location it appeared in the 2009 COS spectrum).
Weak absorption extends all the way down to $\sim1545$ \AA, past the complex
of foreground \ion{C}{4} absorption lines from the Milky Way and the
Magellanic stream.

The spectrum and the best-fit emission model for the Ly$\alpha$ region is
shown in Figure \ref{fig_c4fit}.
Unfortunately, absorption corresponding to \ion{N}{5} falls directly on the
peak and the red side of the steep Ly$\alpha$ emission profile.
Our model of the Ly$\alpha$ emission is perfectly acceptable without any
\ion{N}{5} absorption.
Alternatively, if we include \ion{N}{5} absorption as Gaussian absorption
features approximating the shape of the \ion{C}{4} absorption shown
in Figure \ref{fig_c4fit},
we can obtain an equally good fit. The parameters of the emission line need
only adjust slightly to accommodate the forced inclusion of the absorption
features, so it is impossible to get a reliable
assessment of any \ion{N}{5} absorption, although it is likely present.


There is no absorption visible near \ion{Si}{4}, nor near the lower ionization
transitions of \ion{C}{3}* $\lambda1176$ or \ion{C}{2} $\lambda1334$.
\ion{Si}{4} absorption is visible in the obscuration events seen in
NGC 5548 \citep{Kaastra14}, NGC 985 \citep{Ebrero16},
and NGC 3783 \citep{Mehdipour17}, but it was not present in Mrk 335 in 2009.
Broad \ion{C}{3}* $\lambda1176$ and \ion{C}{2} $\lambda1334$ absorption
(and other lower-ionization species) were present in NGC 5548,
and broad \ion{C}{3}* $\lambda1176$ absorption was present in NGC 985,
Given this wide array of ionization states seen in the obscured states of
other AGN, we measured upper limits for these lines in our current spectrum
of Mrk 335 to provide quantitative constraints on the ionization state of the
absorbing gas.

\subsubsection{Interstellar absorption in {\it FUSE} data}

We now turn to the analysis of the Ly$\beta$+\ion{O}{6} region in the G140L
spectrum.
The rich foreground ISM absorption at wavelengths shortward of 1100 \AA\ makes
this region difficult to model, particularly given the much lower resolution
of the G140L grating. However, as a guide we have retrieved a prior high
S/N spectrum of Mrk 335 from the Mikulski Archive for Space Telescopes
(MAST) obtained in 1999 with the
{\it Far Ultraviolet Spectroscopic Explorer} ({\it FUSE}).
This makes the modeling more tractable since the {\it FUSE} spectrum
can provide a very accurate model for all the foreground absorption.

We start by fitting an emission model to the {\it FUSE} spectrum of Mrk 335
using our fit to the \ion{C}{4} profile as a guide.
We add components for \ion{C}{3} $\lambda$977,
\ion{N}{3} $\lambda$991, the \ion{S}{4} $\lambda\lambda1064,1072$ doublets
beyond \ion{O}{6}, and the \ion{He}{2}/\ion{N}{2} $\lambda$1085 blend.
The continuum power law index is fixed at the shape of the COS G130M+G160M
spectrum (from 1135 to 1800 \AA), $\alpha=1.31$.
We model the foreground ISM absorption using the
{\it FUSE} spectral simulation tool {\tt fsim}
(V5.0, W. R. Oegerle \& E. M.  Murphy).
This model matches the location of every single absorption feature in the
FUSE spectrum, but it is not completely correct for all the line strengths.

This is convincing evidence that there was no prior intrinsic absorption
visible in these earlier high-resolution spectra of Mrk 335. Remarkably, the FUSE data of 1999 provide the ultraviolet view of Mrk~335 nuclear emission  during the X-ray bright state, prior to  2007. 


This ISM model allows us to fit 
the emission spectrum (lines+continuum) of Mrk 335 as
observed with {\it FUSE} very well.
We then use this fit to produce a normalized spectrum of Mrk 335.
Since all emission is described by our model, and we have identified all
absorption features with foreground ISM features (metal lines plus $\rm H_2$),
this normalized spectrum represents the transmission spectrum of the ISM
along the line of sight to Mrk 335.. All features in this spectrum
are foreground ISM absorption, and an independent model of the ISM absorption
(which was not quantitatively accurate) is no longer required.
We then convolve this transmission spectrum with the COS G140L line spread
function to produce a model of the ISM transmission as it would appear in the
COS spectrum.

\subsubsection{Interstellar absorption in COS data}
\begin{figure}[t]
\centering
\resizebox{1.0\hsize}{!}{\includegraphics[angle=270]{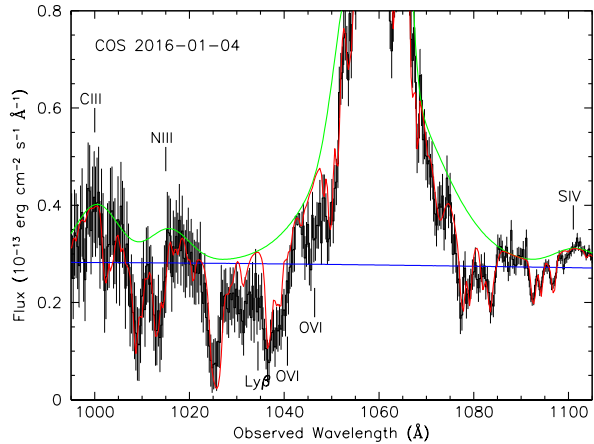}}
\caption{
Best fit to the COS G140L 2016 spectrum of the \ion{O}{6} region
binned by 4 pixels (black histogram).
The peak of the \ion{O}{6} emission line has been truncated to show more detail
at the flux levels of the blue-shifted broad absorption in
\ion{O}{6} and Ly$\beta$.
The solid green line shows the total emission model.
The solid red line tracing the data is the total emission model modified by
our model of the foreground interstellar absorption.
The continuum component is the solid blue line.
Emission lines in Mrk 335 and foreground interstellar absorption lines
are marked.
The intrinsic broad absorption troughs in \ion{O}{6} and Ly$\beta$ are
marked as the depressions below the total model on the extreme blue wing of the
\ion{O}{6} emission line.
}
\label{fig_o6fit}
\end{figure}
Given this model for the complex, contaminating foreground absorption
we can now fit an emission model to the COS G140L spectrum of Mrk 335.
To test for the presence of \ion{O}{6} and Ly$\beta$ absorption,
we first fit the entire region with just emission components.
The weak lines of \ion{C}{3} $\lambda$977, \ion{N}{3} $\lambda$991, and
\ion{S}{4} $\lambda1072$ all require only a single Gaussian component.
For \ion{O}{6}, we require the narrow, semi-broad, and broad components for
each member of the doublet as in \ion{C}{4}, but no very broad component is
necessary.
We constrain the FWHM of the two broad components of the \ion{O}{6}
emission-line profile to match those of \ion{C}{4}, but allow 
the strengths and positions to vary freely.
Our best fit gives $\chi^2=541.51$ for 400 points and 14 free parameters.
This fit is illustrated in Figure \ref{fig_o6fit}.
All narrow absorption features, which correspond to foreground Galactic ISM
absorption, are well matched except in the regions we expect to be affected
by broad Ly$\beta$ and \ion{O}{6} absorption.

One can see that Ly$\beta$ and \ion{O}{6} absorption appear to be present
in our spectrum, but we test for this more quantitatively by
adding in broad Gaussian absorption components for Ly$\beta$ and
\ion{O}{6} based on the locations and shapes of the Ly$\alpha$
and \ion{C}{4} troughs.
We test the significance of adding these components individually
and severally, as summarized in Table \ref{ChisqTab}.
Including absorption components for all three lines, Ly$\beta$,
\ion{O}{6} $\lambda1031$ and \ion{O}{6} $\lambda1037$ gives an improvement
in $\chi^2$ of $\Delta \chi^2 = 32.19$.
For an $F$ test with 3 added free parameters, this is a significant
improvement at greater than 99.9\% confidence.

\begin{table}
        \centering
        \caption{\footnotesize{Statistics of Fits to the Ly$\beta$+\ion{O}{6} Region of Mrk 335}}
        \label{ChisqTab}
\footnotesize
\begin{tabular}{l c c c c}
\hline\hline
Model & $\chi^2$ & $\Delta \rm N_{par}$ & $\Delta\chi^2 / \chi^2_{\nu}$ & P($\Delta\chi^2$) \\
      &          &                      &                & \\
\hline
\hline
No absorption          &  541.51 & 0 & $\phantom{0}0.0\phantom{0}$ & $\ldots$ \\
Add Ly$\beta$ only     &  518.75 & 1 & 16.22 & $<0.001$ \\
Add \ion{O}{6} only     &  532.58 & 2 & $\phantom{0}$6.36 & $0.002$ \\
Add Ly$\beta$+\ion{O}{6} &  509.32 & 3 & 22.95 & $<0.001$ \\
\hline
\end{tabular}
\end{table}

\begin{figure}[!h]
  {\includegraphics[width=1.\columnwidth,angle=0]{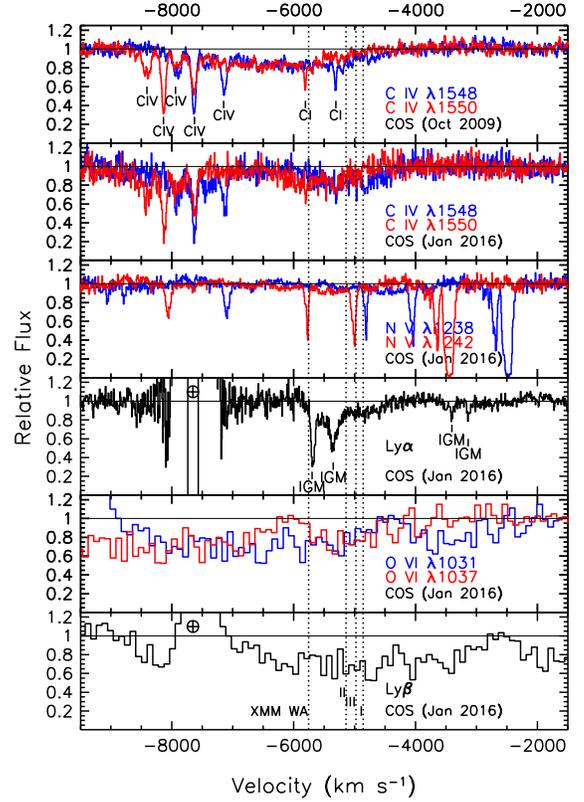}}
  \caption{
Normalized spectra of Mrk 335 showing relative fluxes vs. velocity relative to
the systemic redshift of $z=0.025785$ \citep{Huchra99}.
Top panel: COS spectrum of the \ion{C}{4} region from 2009-10-31.
Second panel: COS spectrum  from this campaign, 2016-01-04.
Red curves give velocities relative to the red component of the doublet, and
blue curves are relative to the blue component of the doublet.
From third to bottom panel: Ly$\alpha$, \ion{O}{6} and Ly$\beta$ regions. 
Dotted vertical lines indicate X-ray absorption velocities (from this work and from \cite{Longinotti13}).
The narrow absorption features in each panel are foreground Galactic
or Intergalactic (IGM) absorption lines. 
}\label{fig:normspectra}
\end{figure}

With our emission-model fits to the Mrk 335 spectra, we can now divide these
into the data to generate normalized spectra.
Figure \ref{fig:normspectra} shows normalized spectra in the \ion{C}{4},
Ly$\alpha$, and \ion{O}{6} regions of the Mrk 335 spectrum illustrating the
broad absorption features.
We also show the (coincident) locations of the X-ray absorption
components identified in \cite{Longinotti13} and in the present paper.

\begin{table}
  \centering
        \caption{\footnotesize{Wavelength Intervals for Mrk 335 Broad Absorption Features}}
        \label{AbsIntervals}
        \footnotesize
\begin{tabular}{l c c c c c}
\hline\hline
Feature & $\lambda_{vac}$ & $v_1$ & $v_2$ & $\lambda_1$ & $\lambda_2$ \\
        &     (\AA)   &  $\rm km~s^{-1}$) & ($\rm km~s^{-1}$) & (\AA) & (\AA) \\
\hline
Ly$\beta$ & 1025.72  & $-5976$ & $-4353$ & 1031.4 & 1037.0 \\
\ion{O}{6} & 1031.93 & $-5989$ & $-4375$ & 1037.6 & 1043.2 \\
\ion{O}{6} & 1037.62 & $-5966$ & $-4648$ & 1043.4 & 1048.0 \\
Ly$\alpha$ & 1215.67 & $-5976$ & $-4362$ & 1222.4 & 1229.0 \\
\ion{C}{4} & 1549.05 & $-8977$ & $-4542$ & 1542.1 & 1565.1 \\ 
\hline
\end{tabular}
\end{table}


\begin{table*}
  \caption{Properties of the UV Broad Absorption Lines in Mrk 335. Velocities and FWHM of O VI and Ly$\beta$ lines are tied to those of C IV, whereas for \ion{N}{5} they are tied to Ly$\alpha$ values. Upper limits of C III*, C II, and Si IV were determined with tabulated values.}
  \label{AbsTab}
\begin{center}
\begin{tabular}{l c c c c c c }
\hline\hline
{Line} & $\lambda_o$ & {EW} & {Velocity} & {FWHM} & $\rm C_f$ & log $\rm N_{ion}$ \\
      & (\AA)       & (\AA) & ($\rm km~s^{-1}$) & ($\rm km~s^{-1}$) & & (log $\rm cm^{-2}$) \\
\hline
\hline
Ly$\beta$ & 1025.72  & $0.91 \pm 0.27$ & $-5\,245$ & $807$ & $0.29 \pm 0.08$ & $>15.43$  \\
\ion{O}{6} & 1031.93  & $0.97 \pm 0.18$ & $-5\,245$ & $807$ & $0.29 \pm 0.08$ & $>14.93$ \\
\ion{O}{6} & 1037.62  & $0.93 \pm 0.12$ & $-5\,245$ & $807$ & $0.27 \pm 0.05$ & $>15.20$ \\
Ly$\alpha$ & 1215.67  & $0.74 \pm 0.03$ & $-5\,144 \pm 36$ & $685 \pm 74$ & $0.22 \pm 0.01$ & $>14.15$  \\
\ion{N}{5}    & 1240.81 & $0.08 \pm0.02$   & $-5\,144 \pm 36$ & $685 \pm 74$  & $0.05 \pm0.03$ &   $>13.86$ \\
\ion{C}{3}* & 1176.01  & $<0.015$ & $-5\,144$ & $700$ & $\dots$ & $<12.15$  \\
\ion{C}{2} & 1334.53  & $<0.015$ & $-5\,144$ & $700$ & $\dots$ & $<12.84$ \\
\ion{Si}{4} & 1393.76  & $<0.017$ & $-5\,144$ & $700$ & $\ldots$ & $<12.28$ \\
\ion{C}{4} & 1549.05  & $2.17 \pm 0.10$ & $-5\,245 \pm 16$ & $807 \pm 75$ & $0.19 \pm 0.04$ & $>14.97$ \\
\hline
\end{tabular}
\end{center}
\end{table*}

We measured the strengths of the absorption features in our spectra of Mrk 335
by directly integrating the normalized spectra shown in
Figure \ref{fig:normspectra}.
We chose velocity intervals covering the full range of absorption visible
by inspection. Note that this interval is substantially larger for the
blended \ion{C}{4} doublet, both because the two components are separated by
$950~\rm km~s^{-1}$ in velocity space, but also because there is no
confusion on the blue wing of the absorber.
The corresponding blue end in Ly$\alpha$ is buried under the damped
Ly$\alpha$ absorption of the Milky Way.
Likewise,  Ly$\beta$ is contaminated by geocoronal Ly$\beta$ emission.
For \ion{O}{6}, the doublet separation is $1814~\rm km~s^{-1}$,
causing blending at higher blue-shifted velocities.
We therefore limited the range for integration to approximately the same
interval used for Ly$\alpha$.
All features have broad widths that are well resolved.
This enables us to directly integrate the normalized absorption profiles to
obtain equivalent widths (EW).
Our direct integrations also yield
column densities using the apparent optical depth method of \cite{Savage91}.
Note that Figure \ref{fig:normspectra} shows that all the troughs have similar
depths. Thus, they are likely saturated, and therefore we quote the column
densities as lower limits.
Given this likelihood of saturation, we also tabulate for each line
the covering fraction implied if the deepest part of the trough is saturated.
The properties of the absorbers are summarized in Table \ref{AbsTab}.

\section{Discussion}
\label{sec:discussion}
The latest X-ray-UV observational campaign carried out  on Mrk~335 in 2015 and presented in this work  fully confirms the scenario proposed by  \cite{Longinotti13} based on non-simultaneous data. We briefly recall the properties of the absorber in Mrk~335  reported therein.
\subsection{The warm absorber in 2009 and 2015} 
 In our comparison with present data we consider only the 2009 spectrum, which in  \cite{Longinotti13} is referred to as the ``the mid-state" and in which the warm absorber properties could be measured at best compared to other epochs. The ionization was there described in terms of the ionization parameter $\xi$=L/nr$^2$  therefore we now quote the corresponding number in terms of U to ease comparison with Table~\ref{tab1}.

Table~\ref{tab2009} reports the values of the three layers of ionized absorption detected in 2009 adapted from Table~5 in  \cite{Longinotti13}. 
The ionized absorber in 2015 (see Table~\ref{tab1}) seems very consistent  with WA~II  in 2009. 
This seems to indicate that since its first record in 2009, the ionized wind has become a persistent feature of Mrk~335, which is also supported by the conclusions reached by \cite{Gallo2018} in their long-term study. These authors conclude that the current low flux state observed since 2007 is not driven by changes in the structure  of the inner accretion disc. Rather, they propose that the variability pattern may be explained either  in terms of coronal changes or intervening absorption.  The apparent stability of the outflow supports therefore the latter hypothesis. We note that in the first low state spectrum of 2007  \citep{Grupe2007}, the presence of ionized absorption could not be investigated in detail due the low S/N of the grating data \citep[see][]{Longinotti13}. However, as the presence of the absorber in this epoch could not be excluded either, we speculate that this wind may well have emerged in 2007 when Mrk~335 entered its  prolonged low X-ray flux state. 
In the following we proceed to explore the possible association of the X-ray  absorbers to the UV wind with the advantage provided  by the simultaneity of the two sets of observations in 2015/16.

 \begin{table}[t]
\caption{\footnotesize{Warm absorber properties in the 2009  RGS spectrum   \citep[adapted from][]{Longinotti13}.}}
\footnotesize
\begin{tabular}{c  c  c  c  c}
\hline
 Phase & Log U       &  Log N$_H$     & v$_{out}$   &   v$_{broad}$  \\
            &    -  &  (cm$^{-2}$) &        (km s$^{-1}$) &  (km s$^{-1}$)   \\
    \hline
I          &   0.39$^{+0.04}_{-0.11}$  & 21.34$\pm$0.06  & 4000$^{+180}_{-700}$ & $\le$100  \\
II        &  1.04$\pm$0.03  & 21.63$\pm$0.06  &   5200$^{+190}_{-60}$  & $\le$100  \\ 
III       &  2.05$^{+0.05}_{-0.09}$ & 22.55$\pm$0.15  &   5300$^{+90}_{-100}$  & $\le$100  \\ 
\hline\hline
\end{tabular} 
\label{tab2009}
\end{table}

\subsubsection{X-ray and UV absorbers in 2015: the RGS view}
\label{subsec:discrgs}
\begin{figure}[t]
 \centering
 {\includegraphics[width=1.1\columnwidth,angle=0]{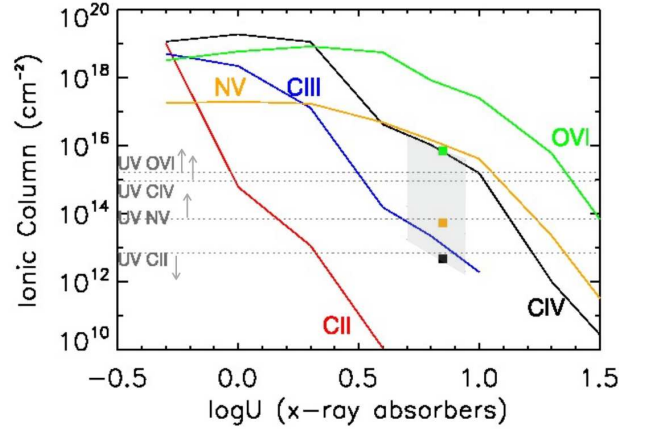}}
  \caption{Comparison of the ionic column densities of UV and X-ray absorbers. Solid lines trace the ionic columns for the X-ray partial covering absorber calculated for the best fit column density of log(N$_H$)=22.99$\pm$0.06. Squared points and grey shade respectively mark the best fit values of the OVI, NV, CIV column densities (same color code) and the area allowed by the X-ray warm absorber detected in RGS (Table~\ref{tab1}). Grey horizontal dotted lines trace the upper and lower limits of the UV lines ionic columns reported in Table~\ref{AbsTab}. }
  \label{fig:ioncol}
\end{figure}

We start by comparing the parameters estimated by the UV lines with those of the soft X-ray warm absorber for which more precise constraints are available owing to the higher detail provided by the grating spectra. 
The outflow velocity in both bands shows remarkable coincidence. 
However, the X-ray-estimated ionic column densities provided by the photoionization model of the soft X-ray absorber in Table~\ref{tab1} are only partially compatible with the columns estimated by the UV throughs (Table~\ref{AbsTab}): this  X-ray warm absorber does not produce enough C IV absorption (N$_{CIV}$=4.7$\times$10$^{12}$~cm$^{-2}$) although it might contribute to the O VI (N$_{OVI}$=7$\times$10$^{15}$~cm$^{-2}$ ) and NV (N$_{NV}$=5.38$\times$10$^{13}$~cm$^{-2}$).
This partial discrepancy can be visualized in Figure~\ref{fig:ioncol} where we can see that the limits traced by the UV columns intersect the X-ray warm absorber columns only for OVI and NV.  

As noted in Section~\ref{sec:rgs}, the spectral fits to the RGS data do not allow us to constrain the width of the X-ray absorption lines, but they seem to favour  the presence of narrow rather than broad absorption lines, which may also poise a problem to interpret the two outflows as arising from the same gas. Nonetheless,  if it is  postulated that X-ray photons cross a smaller range of velocities compared to UV photons as proposed in the sketch of Figure~\ref{fig:wind_sketch}, we may  explain why broader absorption lines are detected in the UV band compared to the narrow lines observed in the X-rays.

Another considerable difference of the wind in the two bands is the covering fraction of the gas. While the coverage of the soft X-ray ionized gas estimated from the RGS is 100\%, the UV absorber covers a small fraction of the ionizing continuum (20-30\%. see Table~\ref{AbsTab}).   
We tested for the presence of a partially covering warm absorber  by using a {\tt PHASE} version with variable covering factor. This test indicates that the warm absorber coverage is consistent with being as low as 80\%, although partial covering is not formally required by the fit statistics. We note that such value is fully coincident with the  constraint obtained for the warm absorber in the mid state flux of 2009.
This may indicate that the hotter part of the outflow (seen in X-rays) is organized in a denser/clumpier structure than the gas ionized by the UV continuum, and/or it may also indicate that the UV source is more extended than the X-ray source, as depicted in Figure~\ref{fig:wind_sketch}. This interpretation is compatible and it may actually explain the presence of the partial covering absorber observed in the broadband X-ray spectrum (see next section).

\subsubsection{X-ray and UV absorbers in 2015: the EPIC-pn view}
The 2015 CCD spectrum of Mrk~335 reveals the presence of two additional layers of absorption: the highly ionized absorber described in Section~\ref{sub:Fekband} ( v$_{out}$=5,200$^{+700}_{-200}$~km~s$^{-1}$, logU=3.13$^{+0.09}_{-0.59}$, logN$_H$$\ge$23.07) and the partial covering absorber described in Section~\ref{sec:broadband} (log(N$_H$)=22.99$\pm$0.06,  C$_f$=0.79$^{+0.02}_{-0.05}$ and logU$<$1.35).
Whereas the former component is too highly ionized  so as to affect the UV spectrum, we explore a possible UV connection with the latter one.

Despite the lack of more detailed properties inferred on the partially covering gas, the X-ray spectral fits show that it has a moderately low ionization, high column density, and that it covers around 80\% of the X-ray source.  Figure \ref{fig:pcov_u} shows that its ionization (logU$<$1.35) overlaps with that of the RGS warm absorber (logU=0.85$^{+0.09}_{-0.14}$), and so does the covering fraction, as reported in Section~\ref{subsec:discrgs}. 
Unfortunately, the outflow velocity of the partial covering gas could not be measured in CCD data (Sec. \ref{sec:pcov_ref}), therefore it is  difficult to pinpoint a more constrained location.  In the spectral fitting, we have tied its velocity to the one of the less dense warm absorber detected in RGS assuming that both are part of the intervening gas that crossed our line of sight during the {\xmm} observation. This is largely justified by the X-ray history of Mrk 335 which, as reported in Sec. \ref{sec:intro}, did not show intervening ionized absorption prior to the decrease of X-ray flux  \citep{Grupe2007,Grupe08} that since 2007 gave rise to the several X-ray campaigns launched on this source.  

Moreover, the relatively high velocity measured in the X-ray and also in the UV absorbers (~5000-6000 km/s) suggests that the obscuring system is located close to the accretion disk or the inner Broad Line Region, and tend to exclude other possible locations placed farther away (e.g. the inner wall of the torus).  These considerations has led us to associate the partial covering and the warm absorber to the same system. 

\begin{figure}[t]
 \centering
 {\includegraphics[width=1.15\columnwidth,angle=0]{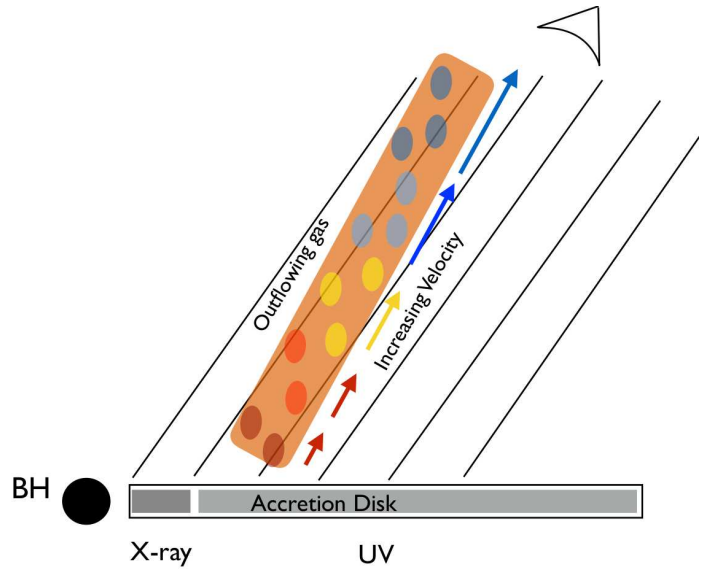}}
  \caption{Sketch of the proposed geometry of Mrk 335  outflow. Black lines represent different sight lines. Due to the different sizes of the sources of X-ray and UV photons, the outflowing gas in the line of sight covers almost entirely the X-ray source and only a small fraction of the UV-emitting disk. The outflowing gas is constituted by a distribution of ionized gas responsible for the X-ray ionized absorber(s) and by denser discrete clouds/filaments that produce the partial covering in the continuum. Colors represent the increasing velocity along the line of sight: X-ray photons cross a small range of velocities compared to UV photons, which may  explain why  UV absorption lines are broader than the narrow lines observed in the X-ray spectrum.}
  \label{fig:wind_sketch}
\end{figure}

We speculate that the two X-ray absorbers detected in the present work with such a wide range of column densities but overlapping ionization state and coverage may well be tracing the same system of gas where denser filaments/clouds are producing the observed spectral curvature in the broadband data whereas less dense parts of the outflow are responsible for imprinting the strong Fe UTA absorption.  A gas with these characteristics is expected to imprint detectable features in the UV spectrum, which are not currently seen. We plot the predicted ionic columns of this absorber in Figure~\ref{fig:ioncol} for the range of the ionization parameter allowed by the best fit. The X-ray curves for the corresponding  UV ions show that a partial covering absorber with logU$\sim$0.5-1 is compatible with the same gas producing both the X-ray and UV absorption. 
As proposed in the previous paragraph, if we postulate that the X-ray and UV absorbers are distributed with very different coverage (80-100\% versus 20-30\%), then we may well explain why the strong X-ray partially covering gas does not appear in the UV data (see Fig.\ref{fig:wind_sketch}).

\subsection{Conclusions}
The obscuring wind in Mrk~335 shows therefore a very rich ionization structure that extends from the UV broad troughs observed by HST up to the highly ionized transitions in the Fe~K band, observed in Mrk~335 for the first time. With exception of the outflow velocity ($\sim$ 5200~km~s$^{-1}$ in CCD data), the properties of this highly ionized wind  (logU=3.13$^{+0.09}_{-0.59}$ and logN$_H$$\ge$23.07) are reminiscent of ultra-fast-outflows that are seen in Seyfert Galaxies \citep{Tombesi11} and whose appearance seems to bear relation with low luminosity states of the sources \citep{Matzeu2017}. The presence of an even faster component of the wind in this low flux state of Mrk~335 cannot be assessed as the EPIC-pn data are heavily affected by high background at E$\ge$8~keV \citep[but see][for the analysis of the flaring portion of this data]{Gallo2019}.
However, we do not exclude the presence of additional and possibly faster outflow components that may also explain unidentified absorption lines
in the RGS spectrum (see Figure~\ref{fig:rgs_oxy}).

 \cite{Longinotti13} extensively discussed the possible interpretation for the appearance of the wind and based on the variability of the broad UV absorption troughs,  concluded that the outflow was transiting our line of sight to the central source at the scale of the Broad Line Region (0.7-4$\times$10$^{16}$~cm). The present data not only brings a strong evidence on the persistency of the wind, as discussed above, but it also provides  a more corroborated  association of the UV and X-ray outflows. 
The absorber therefore can be effectively tracing the base of a radiatively driven wind produced by the accretion disc \citep{Proga04} as suggested for sources with similar behavior as NGC~5548 and NGC~3783 \citep{Kaastra14,Mehdipour17}.  In Mrk~335 the situation seems to be akin to NGC~5548 where the obscuring gas covers 70\% of the source and where the obscuration is observed to extend for several years compared to the isolated eclipsing event that recently characterized NGC~3783 \citep{Mehdipour17}. 

 Further results on the behavior of the absorbers in Mrk~335 are expected by an ongoing multi-epoch study  (Parker et al. in prep.) that include data from very recent X-ray/UV campaigns launched in 2018 and 2019.

\acknowledgments
 \noindent
{\bf ACKNOWLEGMENTS}

\noindent
Based on proprietary observations obtained with  the NASA/ESA {\it HST} and with {\it XMM-Newton}, an ESA science mission with instruments and contributions directly funded by ESA Member States and NASA. 
Some of the data presented in this paper were obtained from the
Mikulski Archive for Space Telescopes (MAST).
This work was supported by NASA
through a grant for {\it HST} program number 13814
from the Space Telescope Science Institute (STScI), which is
operated by the Association of Universities for Research
in Astronomy, Incorporated, under NASA contract
NAS5-26555.
This research has made use of the NASA/IPAC Extragalactic Database (NED),
which is operated by the
Jet Propulsion Laboratory, California Institute of Technology,
under contract with the National Aeronautics and Space Administration.
YK acknowledges support from  DGAPA-PAIIPIT grant IN106518. 
ALL acknowledges support from CONACyT grant CB-2016-01-286316.



\bibliography{agn}

\end{document}